\documentstyle[12pt,epsfig]{article}
\begin{document} 

\def\BS{e^{+}e^{-} \rightarrow  e^{+} e^{-} }
\def\adp{\left({\alpha\over\pi}\right)}
\def\th{\theta}
\def\div#1#2 { {{#1}\over {#2}} }
\def\Ord#1{${\cal O}(#1$)}
\def\DDelta{{\phantom{BHL}}\Delta} 

\begin{titlepage}
\rightline{ \large{ \bf DFUB 96-21} }
\rightline{ \large{ \bf TP-USL/96/12} }
\rightline{ \large{ \bf April 1997} }
\bigskip
\begin{center}
{\large{\bf BHAGEN95: a Monte Carlo program for Bhabha scattering at LEP1/SLC 
and LEP2 energies.} }
\end{center}
\vspace{.5cm}
\begin{center}
{\bf
M.~Caffo$^{ab}$, 
H.~Czy{\.z}\ $^{c \star}$, 
E.~Remiddi$^{ba}$ \\ } 
\end{center}

\begin{itemize}
\item[$^a$]
             INFN, Sezione di Bologna, I-40126 Bologna, Italy 
\item[$^b$] 
             Dipartimento di Fisica, Universit\`a di Bologna, 
             I-40126 Bologna, Italy 
\item[$^c$] 
             Institute of Physics, University of Silesia, 
             PL-40007 Katowice, Poland 

\end{itemize}

\noindent
E-mail: {\tt caffo@bologna.infn.it \\ 
\hspace*{1.3cm} czyz@usctoux1.cto.us.edu.pl \\ 
\hspace*{1.3cm} remiddi@bologna.infn.it \\ } 

\begin{abstract}
We present the Monte Carlo program BHAGEN95, for calculating 
the cross-section of the Bhabha scattering process at LEP1/SLC and LEP2 
energies, usable with continuity from small to large-angle configurations. 
We discuss some improvements in the event generator BHAGEN94, which is 
now part of the new code. 
In particular the weak and QCD corrections are implemented up to two loops 
for the relevant contributions, 
and the emission of one hard photon is treated exactly.
We have included all the radiative corrections 
which are necessary to obtain, for a typical experimental event selection, 
a precision of 0.1-0.2\% at small-angle. At large-angle we estimate a
precision of 0.5\%, with the exception of the region where the beam energy
is a few GeV above the $Z$ boson resonance, where it is up to 1\%.
A detailed comparison with other codes for both small-angle and large-angle 
Bhabha scattering is performed. 
\end{abstract}

\vfill

\footnoterule
\noindent
$^{\star}${\footnotesize 
     Partly supported by the Polish Committee for Scientific Research
     under grant no 2P03B17708 and \\ USA-Poland Maria Sk{\l}odowska-Curie
     Joint Fund II under grant MEN/NSF-93-145. } 

\end{titlepage}

\section{Introduction} 

Bhabha scattering is measured with remarkable precision at LEP1/SLC 
energies in two kinematical regions identified by the scattering angle: 
at small-angle for the luminosity monitoring and 
at large-angle for the measurements of the $Z$ properties. 
\par
Such results require a continuous effort on the theoretical side to maintain 
the theoretical error on the prediction within a comparable precision and 
to provide a tool for realistic comparisons with the experimental data 
(namely an event generator).
\par
We have already given in ~\cite{NC92} the general description of our approach
to the problem, 
while the features of the first version of our event generator BHAGEN are 
described in ~\cite{IJMP93}, with instructions on how to get the program 
and how to use it. In ~\cite{PL94} we have considered the higher order 
QED corrections, including the most relevant contributions and estimating the 
omitted terms; in ~\cite{PL94} a new version of our event generator, BHAGEN94, 
is also announced. 
\par
Several improvements have been made subsequently to BHAGEN94, in 
order to increase and control its precision, and 
in ~\cite{CERN95-03} some results for small-angle 
Bhabha scattering cross-section are presented with the label BG94-NEW 
and compared with the results from few other available 
computer programs: OLDBIS ~\cite{OLDBIS}, LUMLOG ~\cite{LUMLOG}, 
BHLUMI 2.01~\cite{BHLUMI2} and 40THIEVES ~\cite{40THIEVES}.
The conclusion from the comparisons with the above programs, and also from the
use of the new event generator BHAGEN-1PH, which contains the complete hard 
photon matrix element ~\cite{1PH}, is  that the hard photon part (even in the 
improved version of BHAGEN94) deserves a more accurate treatment to achieve 
the 0.1\% precision, presently aimed at small-angle for luminosity monitoring. 
To obtain such a result a new procedure and the code BHAGEN95 are developed. 
BHAGEN95 (available on request from the authors) is actually a collection of 
three programs and in this paper we discuss
in detail two of them, as the third one is already discussed separately 
~\cite{1PH}.
The current precision for the cross-section is heavily dependent on the chosen 
event selection and for our code can be estimated to be about 0.1\%-0.2\% for 
typical experimental event selection at small-angle for LEP1/SLC and LEP2 
energies.
In references ~\cite{CERN96} and ~\cite{BWG_LETT} our results are presented 
in a variety of different event selections and conditions 
(some of those preliminary results are updated in this discussion),
and compared with the other presently available programs of similar
precision:
BHLUMI 4.03 ~\cite{BHLUMI4},
OLDBIS+LUMLOG ~\cite{OBI+LUMG}, 
SABSPV ~\cite{SABSPV},
NLLBHA ~\cite{NLLBHA}.

The precision at large-angle depends on the beam energy and 
also on the event selection used; in the worst case (beam energy a few GeV 
above the $Z$ peak) can be safely estimated to be about 1\%, 
mainly due to the raise of the incertitude coming from the second order leading 
logarithmic correction in the $s-t$-interference terms, which are still 
unknown and usually are assumed to be the same as in the $s$-channel.
For the other energy ranges at LEP1/SLC, and at LEP2,
one can expect with relatively loose cuts a typical accuracy of 0.5\%,
which can be usefully compared with the other presently available programs, 
in the conditions they are suited for: 
ALIBABA ~\cite{ALIBABA}, 
ZFITTER ~\cite{ZFITTER}, 
TOPAZ0 ~\cite{TOPAZ0}, 
BHAGENE3 ~\cite{BHAGENE}, 
UNIBAB ~\cite{UNIBAB},
BHWIDE ~\cite{BHWIDE},
SABSPV ~\cite{SABSPV}.

\par
This paper goes through the following sections:
implementation of an improved energy distribution for the final photon 
emission in the resummed program BHAGEN94, in section 2;
construction of the BG94-FO code, the \Ord{\alpha} version of the event 
generator BHAGEN94, in section 3;
implementation of necessary weak and QCD corrections in BHAGEN94 up
to two loops, in section 4;
construction of the BHAGEN95 code, in section 5;
comparisons at small-angle, in section 6; 
comparisons at large-angle, in section 7; 
conclusions, in section 8.
\bigskip

\section{ Improved energy distribution for the final photon 
emission in the resummed program BHAGEN94.}

In the same notation as in 
~\cite{NC92}, ~\cite{IJMP93}, ~\cite{PL94} and ~\cite{CERN89},
the integrated cross-section is 
\begin{equation}
\sigma(\BS) = \int\limits_{\Delta\Omega_-} d\Omega_- \quad
{d\tilde\sigma(E_b,\th_-)\over{d\Omega_-}} \ ,
\label{eq:s1}
\end{equation}
where the dressed differential cross-section in our approach is 
\begin{eqnarray}
{d\tilde\sigma(E_b,\th_-)\over{d\Omega_-}} = 
&&\int\limits_{E_{min}}^{E_b} {d E^0_- \over{E_b}} F^0(E_b,E^0_-,\th_-)
  \int\limits_{E_{min}}^{E_b} {d E^0_+ \over{E_b}} F^0(E_b,E^0_+,\th_+) 
\nonumber \\
&&\int\limits_{E_{min}}^{E_-} {d E'_-  \over{E_-}} F'(E_-,E'_-,\th_-)
  \int\limits_{E_{min}}^{E_+} {d E'_+  \over{E_+}} F'(E_+,E'_+,\th_+) 
\label{eq:s2} \\
&&{d\sigma(E^0_+,E^0_-,\th_-)\over{d\Omega_-}} \quad \Theta(cuts) \ , \nonumber 
\end{eqnarray} 
$E_b$ is the beam energy, $\th_{\pm}$ the scattering-angle of the electron 
(positron), $s=4E_b^2$ and $t_{\pm}=-s \sin^2{\th_{\pm}\over2}$.
In the laboratory system the electron (positron) energy after 
initial radiation is $E^0_{\pm}$, after the scattering is $E_{\pm}$, 
after the final radiation is $E'_{\pm}$, and $E_{min}$ is the minimum energy 
for the leptons.
\par
In terms of the variables in the LM-c.m.s. (the c.m.s. after initial radiation 
emitted collinearly to the initial direction, as introduced 
in ~\cite{NC92}), the incoming fermion energy 
$E^*=\sqrt{E^0_- E^0_+}$, the scattering angle $\th^*(E^0_-,E^0_+,\th_-)$ 
and the jacobian $\left( {d\Omega^* \over{d\Omega_-}} \right)$,
the differential cross-section from Feynman diagrams, \Ord{\alpha} complete 
and up to \Ord{\alpha^2} leading logarithm, is written as
\begin{eqnarray}
{d\sigma(E^0_+,E^0_-,\th_-)\over{d\Omega_-}} = 
&& \left( {d\Omega^* \over{d\Omega_-}} \right)
   \sum_{i=1}^{10} {{d\bar\sigma_0^{(i)}(E^*,\th^*)}\over{d\Omega^*}} 
\label{eq:s3} \\
&& \left[ 1+\delta_N^{(i)}(E_b,\th^*) \right]
   \left[ 1 + \tilde C^{(i)}(E^*,\th^*) \right] 
   \left[ 1+\delta_N^{(i)}(E^*,\th^*) \right]   \nonumber \ , 
\end{eqnarray}
where ${{d\bar\sigma_0^{(i)}(E^*,\th^*)}\over{d\Omega^*}}$ are the Born 
differential cross-sections for the various channels with vacuum polarization 
insertions (Dyson resummed as in Eq.(2.6) of ~\cite{NC92}),
$\delta^{(i)}_{N}(E, \th)$ (defined in Eq.(16) of ~\cite{PL94}) 
are the leading logarithmic corrections up to \Ord{\alpha^2}, 
and $\tilde C^{(i)}(E_b,\th_-)$ are the \Ord{\alpha} complete corrections 
after subtraction of the terms already included in the previous corrections.
\par 
The function $\Theta(cuts)$ accounts for the rejection procedure, according to 
the requested cuts on energies and angles.
\par 
As usual,
\begin{eqnarray}
\beta(E,\th) &=& \beta_e(E) +\beta_{int}(\th) \ , \nonumber \\ 
\beta_e(E) &=& 2 \adp \left(\ln {{4E^2}\over{m_e^2}} -1\right) \ , \\
\beta_{int}(\th) &=& 4 \adp \ln \left( \tan {\th \over2} \right) \ ,
\nonumber 
\end{eqnarray} 
with $\beta(E_b,\th_{\pm}) \simeq \beta_e(\sqrt{s}/2)$ at large-angle and 
$\beta(E_b,\th_{\pm})  \simeq \beta_e(\sqrt{-t_{\pm}}/2)$ at small-angle.
\par
The emission functions for initial $(F^0)$ and final $(F')$ emission 
in the notation of ~\cite{PL94}, and using the results of ~\cite{S+J}, are
\begin{eqnarray}
F^0(E_b,E^0_{\pm},\th_{\pm}) &&= D^0(E_b,E^0_{\pm},\th_{\pm})
\Biggl\{
     {1\over2} \left( 1 + \left( {E^0_{\pm}\over{E_b}} \right)^2 \right)
     + A^0_m(E_b,E^0_{\pm})    
\label{eq:s4} \\
    &&+ {{\beta(E_b,\th_{\pm})}\over 8}
     \left[ -{1\over2}
     \left( 1 +3 \left( {E^0_{\pm}\over{E_b}} \right)^2 \right)
     \ln \left( {E^0_{\pm}\over{E_b}} \right)
    - \left(1-{E^0_{\pm}\over{E_b}}\right)^2
     \right] \Biggr\}  \ , \nonumber 
\end{eqnarray}
and
\begin{eqnarray}
F'(E_{\pm},E'_{\pm},\th_{\pm}) &&= D'(E_{\pm},E'_{\pm},\th_{\pm})
\Biggl\{
     {1\over2} \left( 1 + \left( {E'_{\pm}\over{E_{\pm}}} \right)^2 \right)
     + A'_m(E_{\pm},E'_{\pm}) 
\label{eq:s5} \\
    &&+ {{\beta(E_{\pm},\th_{\pm})}\over 8}
     \left[ -{1\over2}
     \left( 1 +3 \left( {E'_{\pm}\over{E_{\pm}}} \right)^2 \right)
     \ln \left( {E'_{\pm}\over{E_{\pm}}} \right)
    - \left(1-{E'_{\pm}\over{E_{\pm}}}\right)^2
     \right] \Biggr\}  \ , \nonumber 
\end{eqnarray}
with
\begin{equation}
A^0_m(E_b,E^0_{\pm}) = {1\over2} {{\left( 1 - {{E^0_{\pm}}\over{E_b}} \right)}^2
 \over {\ln \left({{2E_b}\over{m_e}}\right)^2 - 1 } } \ , \quad
A'_m(E_{\pm},E'_{\pm}) ={1\over2}{{\left(1-{{E'_{\pm}}\over{E_{\pm}}} \right)}^2
 \over {\ln \left({{2E'_{\pm}}\over{m_e}}\right)^2 - 1 } } \ .
\end{equation}
The radiator functions for initial $(D^0)$ and final $(D')$ emission are
\begin{equation}
D^0(E_b,E^0_{\pm},\th_{\pm}) = \div12 \beta(E_b,\th_{\pm})
\left( 1-{E^0_{\pm}\over{E_b}} \right)^{\div12 \beta(E_b,\th_{\pm}) -1} 
\label{eq:s6}
\end{equation}
and
\begin{equation}
D'(E_{\pm},E'_{\pm},\th_{\pm}) = \div12 \beta(E'_{\pm},\th_{\pm})
\left( 1-{E'_{\pm}\over{E_{\pm}}} \right)^{\div12 \beta(E'_{\pm},\th_{\pm}) 
-1} \ ,
\label{eq:s7}
\end{equation}
note that the leading logarithmic corrections, up to \Ord{\alpha^2}, 
have been included in the factors $(1+\delta_N^{(i)}(E,\th))$ of the expression 
in Eq.~(\ref{eq:s3}). 

As the differential cross-section in Eq.~(\ref{eq:s3}) does not depend on 
final lepton energies $E'_{\pm}$, in our previous formulations the 
integrations on these variables were performed analytically, assuming the 
further simplification that in the expression for $D'$ in Eq.~(\ref{eq:s7}) 
the values for $\beta(E'_{\pm},\th_{\pm})$ were taken constant at energy $E^*$,
i.e. $\beta(E^*,\th_{\pm})$. 
\par
The simplification allows the analytical integration 
\begin{eqnarray}
\int\limits_{E_{min}}^{E_-} {d E'_-  \over{E_-}} F'(E_-,E'_-,\th_-) &=&
\Delta_-^{\div12 \beta(E^*,\th^*)} G(E^*,\th^*,\Delta_-) \ , 
\label{eq:s8} \\
\int\limits_{E_{min}}^{E_+} {d E'_+  \over{E_+}} F'(E_+,E'_+,\th_+) &=&
\Delta_+^{\div12 \beta(E^*,\th^*)} G(E^*,\th^*,\Delta_+) \ ,
\label{eq:s9}
\end{eqnarray}
where $\Delta_{\pm} = 1-E_{min}/E_{\pm}$ and the function 
$G(E^*,\th^*,\Delta_{\pm})$ is explicitly given in Eq.~(23) of ~\cite{PL94}. 
\par 
The procedure is rather good for small values of $\Delta_{\pm}$, but the 
extension to larger and more realistic values, corresponding to harder photon 
emission, causes a loss in precision, which justifies the difference with
BHLUMI 2.01~\cite{BHLUMI2} and 40THIEVES ~\cite{40THIEVES} amounting up to 
0.8\%, as reported in ~\cite{CERN95-03} for results under label BG94-OLD.
\par
In the new version of BHAGEN94 the $\beta$ dependence on $E'_{\pm}$ is kept
and the integration (or generation) is done numerically without further 
approximations.
The new values (reported in ~\cite{CERN95-03} under label BG94-NEW) 
reduce the difference with the above mentioned programs to about 0.25\%.
\par
The new approach imply the generation of two more variables for the final 
fermion energies $E'_{\pm}$, for the proper description of the peaks in 
Eq.~(\ref{eq:s7}) the variables $y'_{\pm}$ are introduced
\begin{equation}
y'_{\pm} = 
\left( 1-{E'_{\pm}\over{E_{\pm}}} \right)^{\div12 \beta(E_{\pm},\th_{\pm})}\ .
\end{equation}
To generate within a unit volume, the variables are changed to 
$r'_{\pm} \in (0,1)$
\begin{equation}
y'_{\pm} = r'_{\pm} 
\left( 1-{E_{min}\over{E_{\pm}}} \right)^{\div12 \beta(E_{\pm},\th_{\pm})}\ .
\end{equation}
Note that, even if in the generation procedure the function $\beta$ is still 
taken constant in energy, in the actual calculation the exact value is 
computed and used to determine the cross-section.
\bigskip

\section{ Construction of the BG94-FO generator.}

We have extracted from BHAGEN94 a first-order event generator 
\ ---\  i.e. including only corrections up to \Ord{\alpha} \ ---\ 
that we call BG94-FO, 
keeping as much as possible unchanged the structure of the program.
\par 
That is achieved by analytically expanding in $\alpha$ all the formulae 
and then keeping the \Ord{\alpha} terms only; the separation between soft and 
hard radiation is reintroduced through the small emitted energy fraction cut 
$\epsilon$, whose size has to be of the order of the accepted error, 
typically $\epsilon \le 10^{-4}$. 
To allow the expansion in $\alpha$, the integrations in Eq.~(\ref{eq:s2}) are 
split into 
\begin{equation}
\int\limits_{E_{min}}^{E_b} {d E^0_{\pm}} = 
\int\limits_{E_{min}}^{E_b(1-\epsilon)} {d E^0_{\pm}} 
         +\int\limits_{E_b(1-\epsilon)}^{E_b} {d E^0_{\pm}} \ , 
\end{equation}
and
\begin{equation}
\int\limits_{E_{min}}^{E_{\pm}} {d E'_{\pm}} =
\int\limits_{E_{min}}^{E_{\pm}(1-\epsilon)} {d E'_{\pm}} 
         +\int\limits_{E_{\pm}(1-\epsilon)}^{E_{\pm}} {d E'_{\pm}} \ .
\end{equation}
All the resulting terms, which involve more than one integration in the hard 
photon regions $(E_{min},E_b(1-\epsilon))$ and $(E_{min},E_{\pm}(1-\epsilon))$, 
are omitted, being of higher order.
\par
In the soft photon regions 
$(E_b(1-\epsilon),E_b)$ and $(E_{\pm}(1-\epsilon),E_{\pm})$ 
and for smooth functions, 
$E^0_{\pm}$ can be approximated by $E_b$ and $E'_{\pm}$ by $E_{\pm}$, 
while special care has to be devoted to the peaking behaviour of
the functions $D^0$ and $D'$.
Moreover from the definition of $E^0_{\pm}$ (Eq.~(3.6) of ~\cite{NC92}) 
one has that for 
$E^0_{\pm} \simeq E_b$ also $E_{\pm} \simeq E_b$, and $\th_+ = \th_-$, 
so Eq.~(\ref{eq:s2}) becomes
\begin{eqnarray}
{d\tilde\sigma(E_b,\th_-)\over{d\Omega_-}} &\simeq& 
\int\limits_{E_b(1-\epsilon)}^{E_b} {d E^0_- \over{E_b}} D^0(E_b,E^0_-,\th_-)
\int\limits_{E_b(1-\epsilon)}^{E_b} {d E^0_+ \over{E_b}} D^0(E_b,E^0_+,\th_-) 
\nonumber \\
& &\int\limits_{E_b(1-\epsilon)}^{E_b} {d E'_-  \over{E_b}} D'(E_b,E'_-,\th_-) 
   \int\limits_{E_b(1-\epsilon)}^{E_b} {d E'_+  \over{E_b}} D'(E_b,E'_+,\th_-) 
   {d\sigma(E_b,E_b,\th_-)\over{d\Omega_-}} 
\nonumber \\
&+&\int\limits_{E_b(1-\epsilon)}^{E_b}{d E^0_- \over{E_b}}D^0(E_b,E^0_-,\th_-)
   \int\limits_{E_b(1-\epsilon)}^{E_b}{d E^0_+ \over{E_b}}D^0(E_b,E^0_+,\th_-) 
\nonumber \\
& &\int\limits_{E_{min}}^{E_b(1-\epsilon)} 
                           {d E'_-  \over{E_b}} F'(E_b,E'_-,\th_-) 
   \int\limits_{E_b(1-\epsilon)}^{E_b} {d E'_+  \over{E_b}} D'(E_b,E'_+,\th_-) 
   {d\sigma(E_b,E_b,\th_-)\over{d\Omega_-}} 
\nonumber \\
&+&\int\limits_{E_b(1-\epsilon)}^{E_b}{d E^0_- \over{E_b}}D^0(E_b,E^0_-,\th_-)
   \int\limits_{E_b(1-\epsilon)}^{E_b}{d E^0_+ \over{E_b}}D^0(E_b,E^0_+,\th_-) 
\\
& &\int\limits_{E_b(1-\epsilon)}^{E_b} {d E'_-  \over{E_b}} D'(E_b,E'_-,\th_-) 
   \int\limits_{E_{min}}^{E_b(1-\epsilon)} 
                           {d E'_+  \over{E_b}} F'(E_b,E'_+,\th_-) 
   {d\sigma(E_b,E_b,\th_-)\over{d\Omega_-}} 
\nonumber \\
&+&\int\limits_{E_{min}}^{E_b(1-\epsilon)} 
                           {d E^0_- \over{E_b}} F^0(E_b,E^0_-,\th_-)
\int\limits_{E_b(1-\epsilon)}^{E_b} {d E^0_+ \over{E_b}} D^0(E_b,E^0_+,\th_+) 
\nonumber \\
& &\int\limits_{E_-(1-\epsilon)}^{E_-} {d E'_-  \over{E_-}} D'(E_-,E'_-,\th_-) 
   \int\limits_{E_+(1-\epsilon)}^{E_+} {d E'_+  \over{E_+}} D'(E_+,E'_+,\th_+) 
   {d\sigma(E_b,E^0_-,\th_-)\over{d\Omega_-}} 
\nonumber \\
&+&\int\limits_{E_b(1-\epsilon)}^{E_b}{d E^0_- \over{E_b}}D^0(E_b,E^0_-,\th_-)
   \int\limits_{E_{min}}^{E_b(1-\epsilon)} 
                           {d E^0_+ \over{E_b}} F^0(E_b,E^0_+,\th_+) 
\nonumber \\
& &\int\limits_{E_-(1-\epsilon)}^{E_-} {d E'_-  \over{E_-}} D'(E_-,E'_-,\th_-) 
   \int\limits_{E_+(1-\epsilon)}^{E_+} {d E'_+  \over{E_+}} D'(E_+,E'_+,\th_+) 
   {d\sigma(E^0_+,E_b,\th_-)\over{d\Omega_-}} 
\nonumber \ ,
\end{eqnarray}
where the first term accounts for soft emission only, 
the second and third account for hard emission from final electron or positron, 
the fourth and fifth for hard emission from initial electron or positron.
The functions $F^0$ and $F'$ are now intended at \Ord{\alpha}, that is 
omitting the $\beta$ term in curly braces of 
Eq.~(\ref{eq:s4}) and Eq.~(\ref{eq:s5}).
\par
The analytical integration and subsequent expansion in $\alpha$ can be easily 
done for the initial radiation
\begin{equation}
\int\limits_{E_b(1-\epsilon)}^{E_b} 
{d E^0_{\pm} \over{E_b}} D^0(E_b,E^0_{\pm},\th_{\pm}) = 
\epsilon^{\div12 \beta(E_b,\th_{\pm})} = 
1 + {\div12 \beta(E_b,\th_{\pm})} \ln \epsilon +{\cal O}(\alpha^2) \ ,
\end{equation}
and in the case of final radiation the result is similar, apart from 
corrections of order $\epsilon$ 
\begin{equation}
\int\limits_{E_b(1-\epsilon)}^{E_b} 
{d E'_{\pm} \over{E_b}} D'(E_b,E'_{\pm},\th_{\pm}) = 
\epsilon^{\div12 \beta(E_b,\th_{\pm})} +{\cal O}(\epsilon) = 
1 + {\div12 \beta(E_b,\th_{\pm})} \ln \epsilon +{\cal O}(\alpha^2) 
+{\cal O}(\epsilon) \ .
\end{equation}
In Eq.~(\ref{eq:s3}) the expansion in $\alpha$ is performed and only 
\Ord{\alpha} terms are retained: the vacuum polarization contribution 
$\delta^{(i)}_{VP}$ is defined in Eq.(2.3) of ~\cite{NC92},
$\delta^{(i)}_{N}(E_b)$ is taken up to \Ord{\alpha} from Eq.(16) of 
~\cite{PL94}, and $\tilde C^{(i)}(E_b,\th_-)$, as indicated in
~\cite{PL94}, is the nonleading part of the \Ord{\alpha} correction
$C^{(i)}(E_b,\th_-)$, given explicitly in ~\cite{CERN89}. 
When an \Ord{\alpha} correction from above integrals occurs only the Born 
cross-section is retained, and if no longer necessary the transformation 
to LM-c.m.s. is released. 
\par 
With all that the cross-section at \Ord{\alpha} becomes 
\begin{eqnarray}
&&{d\tilde\sigma(E_b,\th_-)\over{d\Omega_-}} \simeq
   \sum_{i=1}^{10} {d\sigma_0^{(i)}(E_b,\th_-) \over{d\Omega_-}}
   \Biggl\{ 1 +2 \beta(E_b,\th_-) \ln\epsilon
   +\delta^{(i)}_{VP} +2 \delta^{(i)}_{N}(E_b) + \tilde C^{(i)}(E_b,\th_-) 
\nonumber \\
&&+\int_{E_{min}}^{E_b(1-\epsilon)} {d E'_- \over{E_b -E'_-}}
{{\beta(E'_-,\th_-)}\over{4}} 
\left[ 1 +\left({E'_-\over{E_b}}\right)^2 
+{\left(1-{E'_-\over{E_b}}\right)^2 \over{\ln (2E'_-/m_e)^2 -1}} \right] 
\nonumber \\
&&+\int_{E_{min}}^{E_b(1-\epsilon)} {d E'_+ \over{E_b -E'_+}}
{{\beta(E'_+,\th_-)}\over{4}} 
\left[ 1 +\left({E'_+\over{E_b}}\right)^2 
+{\left(1-{E'_+\over{E_b}}\right)^2 \over{\ln (2E'_+/m_e)^2 -1}} \right] 
\Biggr\} \label{eq:fo1} \\
&&+\int_{E_{min}}^{E_b(1-\epsilon)} {d E^0_- \over{E_b -E^0_-}} 
{{\beta(E_b,\th_-)}\over{4}} 
\left[ 1 +\left({E^0_-\over{E_b}}\right)^2 
+{\left(1-{E^0_-\over{E_b}}\right)^2 \over{\ln ({2E_b}/m_e)^2 -1}} \right] 
   {d\sigma_0(E_b,E^0_-,\th_-)\over{d\Omega_-}} 
\nonumber \\
&&+\int_{E_{min}}^{E_b(1-\epsilon)} {d E^0_+ \over{E_b -E^0_+}} 
{{\beta(E_b,\th_+)}\over{4}} 
\left[ 1 +\left({E^0_+\over{E_b}}\right)^2 
+{\left(1-{E^0_+\over{E_b}}\right)^2 \over{\ln ({2E_b}/m_e)^2 -1}} \right] 
   {d\sigma_0(E_b,E^0_+,\th_-)\over{d\Omega_-}} 
\nonumber \ ,
\end{eqnarray}
where 
\begin{equation}
{d\sigma_0(E_b,E^0_{\pm},\th_-)\over{d\Omega_-}} =
\left( {d\Omega^* \over{d\Omega_-}} \right)
   \sum_{i=1}^{10} {d\sigma_0^{(i)}(E^*,\th^*) \over{d\Omega^*}}
\end{equation}
with accordingly $E^*=\sqrt{E_b E^0_{\pm}}$ and 
$\th^*=\th^*(E_b,E^0_{\pm},\th_-)$. 
\par
In Eq.~(\ref{eq:fo1}) the last two lines are due to hard initial emission, 
the second and third lines are due to hard final emission; in the approximation 
of constant $\beta$ and symmetrical cuts, the hard final emission contributions 
can be integrated analytically giving 
\begin{equation}
 + \beta(E_b,\th_-) \left[ \ln{\Delta\over{\epsilon}} -\Delta
 +{\Delta^2 \over{4}} {\ln(s/m_e^2) \over{(\ln(s/m_e^2) -1)}} \right] 
   \sum_{i=1}^{10} {d\sigma_0^{(i)}(E_b,\th_-) \over{d\Omega_-}} \ ,
\label{eq:fo2} \end{equation}
with $\Delta = 1 -E_{min}/E_b$.
In Table 1 of ~\cite{CERN95-03} Eq.~(\ref{eq:fo2}) is used to produce the 
results presented under the label BG94-FO-OLD, which 
differ from the exact ones up to 0.87\%, while the direct generation from 
Eq.~(\ref{eq:fo1}) gives the results under the label BG94-FO-NEW, which have 
only up to 0.28\% difference from the exact ones; these are under label OLDBIS 
~\cite{OLDBIS} and BG94-FO-EXACT (obtained using BHAGEN-1PH ~\cite{1PH} for
the hard photon emission part) and are well in agreement inside the OLDBIS 
technical precision of 0.02\%. 
\bigskip

\section{ Implementation of weak and QCD corrections.}

\def\vepae{(v_e^2 + a_e^2)}
\def\eztp{\delta^{tZ}_{1}(+)}
\def\eztm{\delta^{tZ}_{1}(-)}
\def\[(2s^2-1)]{\left(2 \ s_W^2 -1 \right)}
\def\opz{{\left(1+z\right)}^2}
\def\ezt{\delta^{tZ}_{3}(+)}
\def\CHISR{\hbox{Re}\left(\chi(s)\right)}
\def\CHISI{\hbox{Im}\left(\chi(s)\right)}
\def\ezs{\delta^{sZ}_{2}(+)}
\def\ezsp{\delta^{sZ}_{1}(+)}
\def\ezsm{\delta^{sZ}_{1}(-)}

The previous versions of the program ~\cite{NC92}, ~\cite{IJMP93}, including 
the latest BHAGEN94 ~\cite{PL94}, did not contain the $Zee$ weak vertex 
correction, while the self energy corrections were included in an 
approximate form.
For an accuracy better than 1\% at large-angle those complete corrections 
~\cite{HOLLIK}, and also some higher order corrections ~\cite{CERN95-03_2}
are necessary. 
After their inclusion the error due to the neglected weak one loop corrections 
is about 0.06\% around the $Z$ boson resonance (LEP1) and 0.1\% at LEP2 
energies, so smaller than the other left over photonic corrections.
The use of the program at much higher energies (say 1 TeV) requires an update 
of the weak library for having a precision better than 2\%.

For the one loop $Zee$ vertex corrections and  self-energy corrections 
we use the results of ~\cite{HOLLIK}, but we introduce the 
Dyson resummation of the self energy, as described for $s$-channel in
~\cite{CERN89_1} (we do actually the Dyson resummation in all channels).
The higher orders corrections are included as outlined in ~\cite{CERN95-03_2},
using the available results for weak ~\cite{FLEISCHER} 
and QCD corrections ~\cite{Kniehl}.

For completeness we report here the formulae of these corrections as they come
using our notation as in ~\cite{NC92}, ~\cite{IJMP93}, ~\cite{PL94} and 
~\cite{CERN89}.

The $\tilde C^{(i)}(E^*,\theta^*)$ for $(i=4,...,10)$ should be changed in 
the following way

\begin{equation}
   \tilde C^{(i)}(E^*,\theta^*) \rightarrow \tilde C^{(i)}(E^*,\theta^*)
   + C^{(i)}_{Z}(E^*,\theta^*) \ . 
\end{equation}

The $C^{(i)}_{Z}(E^*,\theta^*)$ are 

\begin{equation}
    C^{(4)}_Z ={
        {   8  \eztp s_W^4 + 2  \eztm \[(2s^2-1)]^2 }
           \over {\vepae}} \ , 
\label{eq:c4}
\end{equation}

\begin{eqnarray}
   C^{(5)}_Z &=& {1 \over {\vepae \left( \opz + 4(r_V-r_A)\right) }}
           {\Bigl[ 8  \eztp \opz s_W^4} \label{eq:c5} \\
          & &{ + 2  \eztm \opz \[(2s^2-1)]^2  + 32  \ezt s_W^2
      \[(2s^2-1)] \Bigr]} \ , \nonumber 
\end{eqnarray}

\begin{eqnarray}
   C^{(6)}_Z &=& { 1 \over
    {\vepae^2 \left[\opz (1+4r_V r_A) + 4 (1-4r_V r_A) \right]}}
        { \Bigl[ 256 \eztp \opz s_W^8} \nonumber\\ 
         & &{+ 16 \eztm \opz \[(2s^2-1)]^4 
            + 512 \ezt s_W^4
         \[(2s^2-1)]^2 \Bigr] } \ , 
\label{eq:c6}
\end{eqnarray}

\begin{eqnarray}
    C^{(7)}_Z &=&{ {1} \over { 2 \vepae \Bigl( (1+z^2)r_V + 2 z r_A \Bigr) } }
      { \Bigl[ \Bigl( 8 (1-z^2) \hbox{Re}(\ezs) s_W^2 \[(2s^2-1)]} \nonumber\\ 
    & &{ + 8\opz \hbox{Re}(\ezsp) s_W^4 
          + 2\opz \hbox{Re}(\ezsm) \[(2s^2-1)]^2 \Bigr) } \label{eq:c7} \\ 
    & &{ +   {{\CHISI}\over{\CHISR}}
           \Bigl(  - 8 (1-z^2) \hbox{Im}(\ezs) s_W^2 \[(2s^2-1)]
          - 8 \opz \hbox{Im}(\ezsp) s_W^4 } \nonumber\\ 
    & &{  - 2\opz \hbox{Im}(\ezsm) \[(2s^2-1)]^2 \Bigr) \Bigr] } \ , \nonumber 
\end{eqnarray}

\begin{eqnarray}
    C^{(8)}_Z &=& { 1 \over {\vepae} }
        {  \Bigl[ 8 \hbox{Re}(\ezsp) s_W^4
                         + 2 \hbox{Re}(\ezsm) \[(2s^2-1)]^2 } \label{eq:c8} \\ 
        & &{+ {{\CHISI}\over{\CHISR}}  \Bigl(  - 8 \hbox{Im}(\ezsp) s_W^4
        - 2 \hbox{Im}(\ezsm)\[(2s^2-1)]^2 \Bigr) \Bigr] } \ , \nonumber 
\end{eqnarray}

\begin{eqnarray}
   C^{(9)}_Z &=& {1 \over { {\vepae^2 } \left(1+4 r_V r_A \right)} }
       {   \Bigl[   128 \eztp s_W^8
           + 8 \eztm \[(2s^2-1)]^4} \nonumber\\ 
        & &{ + 128 \hbox{Re}(\ezsp) s_W^8
           + 8 \hbox{Re}(\ezsm) \[(2s^2-1)]^4 } \label{eq:c9} \\ 
        & &{ -{{\CHISI}\over{\CHISR}}  \Bigl( 128 \hbox{Im}(\ezsp) s_W^8 
          + 8 \hbox{Im}(\ezsm) \[(2s^2-1)]^4 \Bigr) \Bigr] } \ ,\nonumber  
\end{eqnarray}

\begin{eqnarray}
  C^{(10)}_Z &=& { 1 \over { (1+z^2) \vepae^2 + 8 z v_{e}^{2} a_{e}^{2} } }
        {\Bigl[ 128 \opz \hbox{Re}(\ezsp) s_W^8 } \label{eq:c10} \\ 
    & &{+ 8 \opz \hbox{Re}(\ezsm) \[(2s^2-1)]^4 
    + 64 (1-z)^2 \hbox{Re}(\ezs) s_W^4 \[(2s^2-1)]^2 \Bigr] } \ .\nonumber  
\end{eqnarray}

The symbols used are 
\begin{eqnarray}
&&\delta_{1}^{tZ}(\pm)=\delta_{1,w,V}^{tZ}(\pm) 
   + {{2}\over{g^Z_{\pm}}} \Pi^{\gamma Z}(t) \ ,
  \delta_{1}^{sZ}(\pm)=\delta_{1,w,V}^{sZ}(\pm) 
   + {{2}\over{g^Z_{\pm}}} \Pi^{\gamma Z}(s) \ , \label{eq:cdef} \\
&&\delta_{2}^{sZ}(\pm)=\delta_{2,w,V}^{sZ}(\pm) 
+ \left( {{1}\over{g^Z_{+}}} + {{1}\over{g^Z_{-}}}\right) \Pi^{\gamma Z}(s) \ ,
  \delta_{3}^{tZ}(\pm)=\delta_{3,w,V}^{tZ}(\pm) 
+ \left({{1}\over{g^Z_{+}}} + {{1}\over{g^Z_{-}}}\right) \Pi^{\gamma Z}(t) \ ,
\nonumber  \end{eqnarray}
while functions $\delta_{i,w,V}^{tZ}(\pm)$, $\delta_{i,w,V}^{sZ}(\pm)$
 ($i=1,2,3$), $\Pi^{\gamma Z}(t)$ and constants $g^Z_{\pm}$
are defined in the second reference of ~\cite{HOLLIK} in the same notation. 
As usual we take 
\begin{equation}
 {\sin^2(\theta_W) \equiv s_W^2 = 1 - c_W^2  = 1-{{M_W^2}\over{M_Z^2}} } \ .
\label{eq:c11} 
\end{equation}

The above formulae take into account the $Zee$ weak vertex corrections and 
$\gamma Z$ mixing self energy contribution. 
To include completely the $Z$ boson self energy corrections it is sufficient 
to change in our approach the two functions $\chi(s)$ and $\chi'(t)$. 
The new ones are 
\begin{eqnarray}
  \chi(s) &=& { {{1} \over {16 s_W^2 c_W^2}} {{1}\over{1+\Pi^Z(s)}}
   \vepae {{s}\over{s-M_Z^2 +is\Gamma_Z/M_Z}} }\ ,
           \label{eq:c12} \\  
  \chi'(t) &=& { {{1} \over {32 s_W^2 c_W^2}} {{1}\over{1+\Pi^Z(t)}}
   \vepae {{s}\over{M_Z^2 -t}} } \ ,
           \label{eq:c13} 
\end{eqnarray}
where 
$\Pi^Z(s) = \hbox{Re}[\hat \Sigma^Z(s)]/(s-M_Z^2)$ and $\hat \Sigma^Z(s)$ 
is again defined in ~\cite{HOLLIK}.
All the higher order corrections are included into the 
$\Pi^Z$ and $\Pi^{\gamma Z}$ functions, and through them into the formulae
~(\ref{eq:c4}-\ref{eq:cdef}) and ~(\ref{eq:c12}-\ref{eq:c13}), which remain
unchanged.

The value of $s_W^2$ is calculated using Eq.~(\ref{eq:c11}) with $M_Z$
as an input parameter. In the code the value of $M_W$ can be supplied by 
the user, or is calculated using the Fermi constant $G_F$ as an independent 
parameter.
In the latter case we implement the approach outlined in ~\cite{CERN95-03_2},
which allows to get the value of the $W$ boson mass with higher accuracy
than from its direct measurement.  
For this better precision it is necessary to include the already mentioned 
weak corrections up to two loops to the $\Delta\rho$ parameter 
~\cite{FLEISCHER} and also the QCD corrections to the self energy functions of 
the bosons (which are needed to properly calculate $\Delta r$ ) ~\cite{Kniehl}. 
Checking with the values of the $W$ boson mass presented in ~\cite{CERN95-03_2} 
as a function of the mass of the top quark $m_t$ (at fixed values of 
$m_H=300$ GeV and $\alpha_S(M_Z)=0.125$), we obtain a very good agreement 
around $m_t=175$ GeV (only a small difference (0.03\%) is appreciable 
above the unrealistic value $m_t$ = 225 GeV). 

Actually the inclusion in BHAGEN94 (and as a consequence also in BHAGEN95) of 
the higher order QCD corrections is done in an approximate form. 
In particular for the first two fermion generations (u,d,s and c quarks, 
for which the limit $s,-t \gg m^2$ is valid, with $m$ the quark mass) 
we use the multiplicative factor $(1+{{\alpha_S}\over \pi}+...)$ 
in front of the contribution due to one quark loop in the self energy 
functions of the bosons. 
For the $t,b$ doublet we use the approximated formulae, 
presented in ~\cite{Kniehl}, 
for the doublet contribution to $\Delta r$ in the $M_W$ calculation.
On the contrary in the Bhabha cross-section calculations is necessary the 
knowledge of the separate contributions $\Pi^Z$ and $\Pi^{\gamma Z}$ to the 
boson self energies, we use them in the form described in ~\cite{CERN95-03_2} 
(BHM/WOH approach), i.e. we include the leading $\Delta\rho^{HO}$ corrections. 
In this way the corrections are
calculated in the point $s=M_Z^2$ and their $s,t$ dependance 
(which is known however only in the QCD two loop contribution) is not taken
into account. Also some known nonleading corrections are not included. 
All this considered the approximation is sufficient to reach at large-angle 
the projected accuracy of 0.5\% in the cross-section calculations. 
Moreover other not included QED two loop corrections are expected to be 
much bigger than the left over QCD corrections, like the next to leading 
correction, which is not yet calculated, or the inferred leading term in 
$s-t$ channel interference, assumed to be identical to the $s$-channel.
\bigskip

\section{ The BHAGEN95 code. }

BHAGEN95 is a collection of three programs to calculate the cross-section
for Bhabha scattering from small to large scattering angles at LEP1 and
LEP2 energies.
In its present form the integrated cross-section $\sigma({\hbox{BHAGEN95}})$ 
for a given selection of cuts is calculated as
\begin{eqnarray}
        \sigma({\hbox{BHAGEN95}})
                = \sigma({\hbox{BHAGEN94}})
                -\sigma^H({\hbox{BH94-FO}})
                +\sigma^H({\hbox{BHAGEN-1PH}}) \ .
       \label{eq:cs}
\end{eqnarray}

$\sigma({\hbox{BHAGEN94}})$ is the integrated cross-section, 
based on leading \Ord{\alpha^2} exponentiated formulae,   
obtained with the Monte Carlo event generator BHAGEN94, 
with the implementations described in the previous sections 2 and 4.
The use of collinear kinematics of initial and final radiation leaves 
some approximation in the angular distribution, which limits the accuracy
particularly in the region of hard photon emission, as remarked at the end of
section 2.
\par
$\sigma^H({\hbox{BH94-FO}})$ is the integrated cross-section of
\Ord{\alpha} for one hard photon emission, obtained selecting the appropriate 
part in the Monte Carlo event generator BH94-FO, the \Ord{\alpha} expansion 
of BHAGEN94 described in section 3.
\par
$\sigma^H({\hbox{BHAGEN-1PH}})$ is the integrated cross-section obtained
with the one hard photon complete matrix element and exact kinematics,
implemented in the Monte Carlo event generator BHAGEN-1PH ~\cite{1PH}.
\par
The subtraction of $\sigma^H({\hbox{BH94-FO}})$ and its substitution
with $\sigma^H({\hbox{BHAGEN-1PH}})$ is performed to reduce the error
in the cross-section, coming from the approximation in the contribution 
due to the one hard photon emission in BHAGEN94, as discussed in 
~\cite{CERN95-03}.
In this way the one hard photon emission is treated exactly in BHAGEN95.

The choice of the energy threshold $\epsilon$ (in units of the beam energy) 
for a photon to be considered hard should not affect the difference between 
$\sigma^H({\hbox{BHAGEN-1PH}})$ and $\sigma^H({\hbox{BH94-FO}})$. 
However the approximate hard photon treatment in BH94-FO implies that a 
correction of the order of $\epsilon$ is missing, so it is necessary to 
choose $\epsilon$ well below the desired accuracy, typically in the range 
of $10^{-5}-10^{-4}$.
But even so the difference is really independent on $\epsilon$ only if 
the soft photon treatment in BH94-FO is the same as in the exact program 
BHAGEN-1PH. This is indeed the case.
In fact changing $\epsilon$ from $10^{-4}$ to $10^{-5}$ the separate 
cross-sections are growing about 20-30\% for the event selections described 
in ~\cite{CERN96}, while the difference is stable within statistical errors.

The three programs provide cross-sections, which are summed as in 
Eq.~(\ref{eq:cs}) or used to obtain other quantities, such as 
forward-backward asymmetry.
Although the constituent programs are separately genuine event generators, 
the discussed combination can be used for Monte Carlo integration only.

At small-angle we estimate the accuracy in the cross-section evaluation,
due to the uncontrolled higher orders terms \Ord{\alpha^2 L} and
\Ord{\alpha^3 L^3} and to the incertitude in \Ord{\alpha^2 L^2} $s-t$
interference, to amount comprehensively to about 0.1\%.
The further error, due to the approximate two hard photon contribution 
(strongly dependent on the imposed cuts) is estimated on the basis of the 
calculated correction for the one hard photon contribution times 
$\beta_e(s)\simeq 0.1$, to account for the increase in perturbative order.

In Table \ref{Tab1} are presented the values of the ratio $R$, defined as 
 \begin{eqnarray}
         R  = 100 {{\sigma^H({\hbox{BHAGEN-1PH}})
                -\sigma^H({\hbox{BH94-FO}}) } \over
                 {\sigma({\hbox{BHAGEN95}})}}
                   \ ,
       \label{eq:cs1}
\end{eqnarray}
which is the correction (in \%) introduced in BHAGEN95, to account for the 
approximation in the one hard photon contribution in BHAGEN94. 
The correction is given for the event selections used in ~\cite{CERN96}, 
in the same notation as for the $t$-channel comparison. 
As already said, for $\beta_e(s)\simeq 0.1$ the same numbers in Table 
\ref{Tab1} are an estimation (in per mil) of the inaccuracy, due to the 
approximate treatment of two hard photon emission.
The values in Table \ref{Tab1} are acceptably small for cuts closer to the 
ones in experiments,  which are of the angular asymmetric WN type, with 
$ 0.3 \le z_{min} \le 0.7 $, where $z_{min}$ is the energy threshold for the
final clusters for accepting or rejecting the events. 
Indeed in such cases the (included) one hard photon 
corrections are found to be below 1.4 \% at LEP1, so that the corrections 
to two hard photon emissions are expected to be below 1.4 per mil. 
For the values at $z_{min}=0.9$ and calorimetric event selection, 
these corrections are expected to be much bigger (at a few per mil level).

All included we estimate at small-angle an accuracy of the order of
0.1\%-0.2\% for typical experimental cuts for both LEP1 and LEP2 energies.
\par
At large-angle we estimate the accuracy of the \Ord{\alpha^2 L^2} $s-t$ 
interference contribution up to 1\% (depending on energy and cuts) at LEP1, 
but much smaller at LEP2.
The error, coming from the approximate treatment of two hard photon emission,
is estimated as explained above for the small-angle case, and is smaller for 
more stringent acollinearity cut.
All included we estimate an accuracy of the order of 1\% in the worst case
at LEP1, when the beam energy is a few GeV above the $Z$ boson peak, while
typical accuracy for relatively loose cuts is 0.5\%.

The three programs run separately.
They provide initialization and fiducial volume definition according to
input parameters, then starts the generation of events according to 
appropriate variables, which smooth the cross-section behavior.
Rejection is performed through the routine {\tt TRIGGER}, where the special
cuts can be implemented.
The programs stop when the requested number of accepted events is reached
or alternatively when the requested accuracy is obtained.

The following data have to be provided in input:
mass of the $Z$, mass of the top quark, mass of the Higgs, value of
$\alpha_S(M_Z)$, value of $\Gamma_Z$, the beam energy $E_{beam}$, the
minimum energy for final leptons $E_{min}$ (larger than 1 GeV), 
minimum and maximum angle for the scattered electron (positron) with 
the initial electron (positron) direction, maximum acollinearity allowed 
between final electron and positron, number of accepted events to be produced, 
numbers to initialize the random number generator.
The following possibilities are also provided: i) to switch on or off the 
leading contribution from virtual and soft emitted pairs ~\cite{PL94},
ii) to calculate separately the different channel contributions (useful for 
tests), iii) the recording of the generated events of each component program 
in a separate file.

For \Ord{\alpha} programs the minimum and maximum energy allowed for the photon
has to be specified.
The input of BHAGEN-1PH requires also the maximum acoplanarity, 
and minimum angles of the emitted photon with initial 
and final fermion directions, if the contributions with the collinear photons
are to be excluded.

Each program returns the input parameters and the values of the cross-section
obtained with weighted and unweighted events, with the relative statistical
variance (one standard deviation).
Of course, due to the efficiency, the weighted cross-section is usually much
more precise than the unweighted one.
The total integrated cross-section is then calculated according to Eq.
(\ref{eq:cs}).

\begin{table}[!ht]
\centering
\begin{tabular}                            {|c|c|c|c|c|c|}
\hline
$ z_{min} $
& 0.1
& 0.3
& 0.5
& 0.7
& 0.9
\\
\hline
\hline
  BARE1(WW)  
& 
& -0.24
& -0.29
& -0.23
& -0.11
\\
\hline
  CALO1(WW)
& 
& -0.46
& -0.75
& -1.15
& -2.29
\\
\hline
  CALO2(WW)
& -0.38
& -0.60
& -1.01
& -1.66
& -3.57
\\
\hline
  SICAL2(WW)
& +0.35
& +0.26
& +0.005
& -1.02
& -3.34
\\
\hline
  CALO2(NN)
& -0.54
& -0.73
& -1.06
& -1.76
& -3.66
\\
\hline
  CALO2(NW)
& +0.01
& -0.22
& -0.64
& -1.39
& -3.40
\\
\hline
  CALO2(WW-all incl.)
& -0.36
& -0.57
& -0.96
& -1.59
& -3.42
\\
\hline
  CALO2(NN-all incl.)
& -0.52
& -0.70
& -1.02
& -1.69
& -3.51
\\
\hline
  CALO2(WN-all incl.)
& +0.00
& -0.21
& -0.61
& -1.33
& -3.26
\\
\hline
\hline
  CALO3(NN)
& -1.14
& -1.29
& -1.56
& -2.19
& -4.38
\\
\hline
  CALO3(NW)
& -0.21
& -0.41
& -0.77
& -1.61
& -3.95
\\
\hline
  CALO2(WW)
&  
& -0.61
& -1.02
& -1.68
& -3.62
\\
\hline
  SICAL2(WW)
&  
& +0.25
& +0.04
& -1.03
& -3.40
\\
\hline
\end{tabular}
\caption{\small 
Correction R, defined in Eq.~({\protect\ref{eq:cs1}}), due to the approximate 
treatment of the one hard photon emission in BHAGEN94, in \% of the BHAGEN95 
cross-section. 
All event selections are defined in ~\protect\cite{CERN96} in the same 
notation. 
First 9 rows are for LEP1 and last 4 rows are for LEP2 energies.
The option 'all incl.' means that all possible contributions are included, 
while the remaining tests are done only for $t$-channel contributions with 
vacuum polarization switched off. }
\label{Tab1}
\end{table}
\bigskip

\section{ Comparisons at small-angle. }

The \Ord{\alpha} results are tested with other programs in ~\cite{CERN95-03} 
for bare event selection and in ~\cite{CERN96} also for calorimetric event 
selection, where the precision of 0.03 \% is confirmed, as a consequence 
of the agreement of different programs within statistical errors. 

In ~\cite{CERN95-03} the results of BHAGEN94, including the improvement 
outlined here in section 2, are presented for a comparison of the 
cross-section with the final electron and positron scattering 
angles $\th_\pm$ in the range $3^{\circ} \leq \th_\pm \leq 8^{\circ}$.
From this comparison it is clear that we cannot reach a precision better than
0.3\% using only the structure function approach. 

To have a better precision the code BHAGEN95 is settled, with the features 
described in the previous section. 
Several preliminary results are already presented and compared in 
~\cite{CERN96}, so we correct the few which are revised, but 
we do not repeat here the unchanged ones. 

In Fig. 16 of ~\cite{CERN96}, in the comparison at small-angle 
($t$-channel only) the bare event selection case (BARE1, WW) shows 
good agreement (almost inside the 0.1\% box) 
up to $z_{min} = 0.9$ included among all the programs (BHAGEN95, 
BHLUMI 4.03 ~\cite{BHLUMI4},
OLDBIS+LUMLOG ~\cite{OBI+LUMG}, 
SABSPV ~\cite{SABSPV},
NLLBHA ~\cite{NLLBHA}).

For calorimetric event selections BHAGEN95 and OLDBIS+LUMLOG 
have very close results for every event selection cut, but they 
agree (within 0.1\% at LEP1 and 0.2\% at LEP2) with BHLUMI and SABSPV 
only for values of the parameters close to the real experimental ones 
(NW or WW, $0.3\le z_{min} \le 0.7$).
The difference is larger for more severe cuts (as in the NN case) and  
for $z_{min} = 0.9$. 

It is proposed in ~\cite{CERN96} that for OLDBIS+LUMLOG the difference is due 
to the so called 'classical limit' (i.e. zero radiation emission limit)
which could be different in higher orders.

For BHAGEN95 we believe that the difference is due to the approximate 
treatment of two hard photons illustrated in previous section, while the
'classical limit' is sufficiently accurate.
In fact for calorimetric event selection, even for $z_{min} \rightarrow 1.0$, 
that limit is not approached, as almost collinear hard photons can join the
final lepton in the cluster.
The 'classical limit' is approached only for BARE1 event selection, where
all the programs are in substantial agreement even at $z_{min} =0.9$, 
and remarkably there the estimated error in BHAGEN95 on the basis of 
Table \ref{Tab1} is only about 0.01 \%.
On the contrary for calorimetric event selection and $z_{min} = 0.9$ 
hard photons are included within very stringent cuts on phase space, 
as the cluster opening is very narrow. 
The structure function approach (used in the part of BHAGEN95 called BHAGEN94) 
for this configuration is expected to be inaccurate, as it is not able to mimic 
very sophisticated cuts.

In conclusion, at small-angle for the test cross-section called 
'$t$-channel only', the results of BHAGEN95 are in very good agreement 
with those of OLDBIS+LUMLOG, and inside the errors also with those of 
BHLUMI and SABSPV. 
The accuracy of BHAGEN95 is very much dependent on the event selection used, 
and is of the order of 0.1\% - 0.2\% for typical experimental cuts, 
while for more stringent cuts it can amount up to say 0.5\%, with the 
source of the error well understood.

The results for the complete cross-section of BHAGEN95, with all the other 
contributions included, are presented and compared in Table 18 and in Fig. 17 
of ~\cite{CERN96}, repeated here in Fig. \ref{fig:sical92asy-FG} for 
completeness
\footnote{The values of BHAGEN95 in Fig. 2 of ~\cite{BWG_LETT} differ 
          slightly from the ones in ~\cite{CERN96}, due to a second bug 
          introduced in the rush-fixing of a previous (actually inactive) 
          bug in the routine {\tt TRIGGER}. 
          The corrected results confirm those already presented in 
          ~\cite{CERN96}. }.

It is interesting to compare the difference $\Delta\sigma$ between the 
complete results (all included) of Table 18 and the '$t$-channel only' 
results of Tables 14 and 16 of ~\cite{CERN96} for BHLUMI and BHAGEN95 
(for SABSPV values the comparison has no significance due to larger 
statistical errors). 
In Table \ref{Tab2} are presented the values for different angular ranges 
(WW, NN, WN) and energy threshold $z_{min}$, taken from Table 18 of 
~\cite{CERN96} for BHLUMI and calculated as indicated above for BHAGEN95.

The difference $\Delta\sigma$ is mainly due to vacuum polarization correction, 
presumably implemented in the same way in both programs, using also the 
parameterization in ~\cite{VPH}, and to the $Z-\gamma$ interference term.
This latter contribution at small-angle is in the range of 1 per mil of 
the total cross-section, depending mostly on the angular range allowed and 
much less on the other details of the selection.

In the last column of Table \ref{Tab2} is presented in per mil the variation 
 \begin{eqnarray}
      V = 1000 {{\Delta\sigma({\hbox{BHAGEN95}})
                -\Delta\sigma({\hbox{BHLUMI}}) } \over
                 {\sigma({\hbox{BHLUMI,all incl.}})}}
                   \ ,
       \label{eq:cs2}
\end{eqnarray}
between the complete cross-section obtained with BHLUMI and the one obtained 
with BHAGEN95, due to the difference in vacuum polarization implementation and 
$Z-\gamma$ interference correction. 
It can be seen that in the experimentally interesting region the variation 
is statistically significant, always positive and increasing with $z_{min}$, 
and it is almost up to 0.1\%.
\begin{table}[!ht]
\centering
\begin{tabular}                            {|c|c|c|c|}
\hline
$ z_{min} $
& $\Delta\sigma$(BHLUMI,WW)
& $\Delta\sigma$(BHAGEN95,WW)
& V(WW)
\\
\hline
0.1
&5.140(8) 
&5.203(11)
&0.46(10)
\\
\hline
0.3
&5.126(8)
&5.197(14)
&0.52(12)
\\
\hline
0.5
&5.100(8)
&5.195(14)
&0.70(12)
\\
\hline
0.7
&4.994(8)
&5.125(16)
&0.99(14)
\\
\hline
0.9
&4.627(8)
&4.821(16)
&1.57(15)
\\
\hline
 
& $\Delta\sigma$(BHLUMI,NN)
& $\Delta\sigma$(BHAGEN95,NN)
& V(NN)
\\
\hline
0.1
&3.751(7)
&3.802(14)
&0.51(16)
\\
\hline
0.3
&3.742(7)
&3.801(14)
&0.60(16)
\\
\hline
0.5
&3.728(7)
&3.799(14)
&0.72(16)
\\
\hline
0.7
&3.678(7)
&3.774(14)
&0.99(16)
\\
\hline
0.9
&3.430(7)
&3.579(14)
&1.64(18)
\\
\hline

& $\Delta\sigma$(BHLUMI,WN)
& $\Delta\sigma$(BHAGEN95,WN)
& V(WN)
\\
\hline
0.1
&3.883(4)
&3.920(13)
&0.36(14)
\\
\hline
0.3
&3.873(4)
&3.919(13)
&0.45(14)
\\
\hline
0.5
&3.854(4)
&3.916(13)
&0.61(14)
\\
\hline
0.7
&3.779(4)
&3.869(13)
&0.90(14)
\\
\hline
0.9
&3.478(4)
&3.624(14)
&1.59(15)
\\
\hline
\end{tabular}
\caption{\small 
The difference $\Delta\sigma$, in nb, between the complete cross-section 
and the '$t$-channel only' contribution, for BHLUMI and BHAGEN95, for several
event selections (WW, NN, and WN angular range), for $2E_b = 92.3$ GeV.
In the last column is the variation V, defined in 
Eq.~({\protect\ref{eq:cs2}}),  
giving in per mil the difference due to vacuum polarization implementation and 
$Z-\gamma$ interference correction of BHAGEN95 respect to BHLUMI. 
In brackets is the statistical error on the last digits. } 
\label{Tab2}
\end{table}

It seems difficult that the difference comes from vacuum polarization 
implementation, so we have investigated the $Z-\gamma$ interference 
contribution. 

In the next section are reported comparisons at large-angle: BHAGEN95 is in 
a good agreement (about 0.5\%) with the other programs (notably with 
ALIBABA for BARE event selection), where the contribution of the $Z-\gamma$ 
interference term amounts up to half of the cross-section.

The accuracy of the $Z-\gamma$ interference term included in BHLUMI is tested 
in ~\cite{J+P+W}, relaying mostly on the comparison with the ALIBABA code, 
so we have obtained results in the same conditions for BHAGEN95 for comparison.
The results of ALIBABA and the Born cross-sections, to which the
results are normalized, were taken from ~\cite{B+P92} for old generation 
luminometers (LCAL: larger angular range between $3.3^{\circ}$ and 
$6.3^{\circ}$) and from ~\cite{B+P93} for new generation luminometers 
(SCAL: smaller angular range between $1.5^{\circ}$ and $3.15^{\circ}$, more
similar to the ones used in ~\cite{CERN96}), 
while the results of BHLUMI are taken from ~\cite{J+P+W} in the same 
conditions. 
In Table \ref{Tab3} the results of ALIBABA, BHLUMI and BHAGEN95 
in the SCAL angular range are inside the 0.01\% difference, and in the LCAL 
angular range are inside the 0.05\%.
This test is done for very loose cuts: no cut on photon energy and angles
is applied and only the angular ranges of leptons are restricted.
In the results of Table \ref{Tab2} the agreement between BHLUMI
and BHAGEN95 is much better for the looser cuts, than for the more severe ones,
and this could be the reason for the better agreement in Table \ref{Tab2} .

It is however difficult to identify with certainty the source of the difference 
without a more detailed analysis, particularly if it comes from a different
use of the same electroweak parameters, but it would be necessary in case the 
accuracy has to be better than 0.1\%.
\begin{table}[!ht]
\centering
\begin{tabular}                            {|c|c|c|c|c|c|c|}
\hline
          &
\multicolumn{3}{|c|}{LCAL angular range}
&\multicolumn{3}{c|}{SCAL angular range} \\
\hline
$ 2 E_b $ 
&BHAGEN95 
&ALIBABA
&BHLUMI
&BHAGEN95
&ALIBABA
&BHLUMI
\\
\hline
89.661 
&0.757
&0.778
&0.794
&0.165
&0.172
&0.175
\\
\hline
90.036 
&0.779
&0.799
&0.816
&0.169
&0.177
&0.180
\\
\hline
90.411 
&0.724
&0.747
&0.754
&0.156
&0.164
&0.165
\\
\hline
90.786 
&0.527
&0.545
&0.538
&0.112
&0.117
&0.116
\\
\hline
91.161 
&0.175
&0.187
&0.158
&0.035
&0.036
&0.032
\\
\hline
91.563 
&-0.206
&-0.206
&-0.247
&-0.048
&-0.050
&-0.057
\\
\hline
91.911 
&-0.473
&-0.479
&-0.521
&-0.105
&-0.110
&-0.116
\\
\hline
92.286 
&-0.602
&-0.609
&-0.647
&-0.132
&-0.136
&-0.143
\\
\hline
92.661 
&-0.642
&-0.650
&-0.678
&-0.141
&-0.145
&-0.150
\\
\hline
\end{tabular}
\caption{\small 
$Z-\gamma$ interference term (in \% of Born cross-section)
 for BHAGEN95, ALIBABA and BHLUMI for two
different angular ranges (LCAL and SCAL) as a function of the beam energy. }
\label{Tab3}
\end{table}
\bigskip

\section{ Comparisons at large-angle. }

At large-angle some comparisons were presented for LEP1 and LEP2 energies 
in ~\cite{CERN96}.
However, as mentioned also there, the higher orders weak and QCD corrections 
were just implemented in BHAGEN95 and the results were very preliminary. 
Unfortunately some bugs were present in the program and, after their 
correction and test, the new results of BHAGEN95 can now be compared with 
the ones of the other programs 
ALIBABA ~\cite{ALIBABA}, 
TOPAZ0 ~\cite{TOPAZ0}, 
BHAGENE3 ~\cite{BHAGENE}, 
UNIBAB ~\cite{UNIBAB},
BHWIDE ~\cite{BHWIDE},
SABSPV ~\cite{SABSPV}.

We report here the new results of BHAGEN95 in Fig. \ref{fig:lab-lep1-bare} 
and Fig. \ref{fig:lab-lep1-calo} for LEP1 and in Fig. \ref{fig:lab-lep2-calo} 
for LEP2 energies (corresponding respectively to Fig. 19, 20 and 
Fig. 21 in ~\cite{CERN96} in the same notation), for scattering angles 
$40^\circ < \th_- < 140^\circ$ and $0^\circ < \th_+ < 180^\circ$. 

We estimate that at large-angle the accuracy of the program BHAGEN95 is 0.5\% 
everywhere, except in the region of a few GeV above the $Z$ boson peak, 
where, due to the absence of the calculation of the leading \Ord{\alpha^2 L^2} 
term in $s-t$ channel interference, it can be assumed to be up to 1\% 
~\cite{PL94}.

In Fig. \ref{fig:lab-lep1-bare}, for LEP1 energy and BARE type event selection 
the results of BHAGEN95 are in agreement with most of the programs inside 0.5\% 
for energies under and on top of the $Z$ boson peak, while above the resonance 
and with more stringent acollinearity cuts ($10^\circ$) there is 
a noticeable agreement with ALIBABA, but some difference from other programs, 
which however is still less than 1\%. 
The same pattern is kept for the calorimetric event selection CALO, 
shown in Fig. \ref{fig:lab-lep1-calo}, where unfortunately results from 
ALIBABA are no longer available. 

In Fig. \ref{fig:lab-lep2-calo}, for LEP2 energies and CALO event selection, 
there is a remarkable clustering in a 2\% interval of most of the programs.
\bigskip

\section{ Conclusions.} 

The Monte Carlo integrator BHAGEN95 (available on request from the authors) 
calculates the cross-section of the Bhabha scattering with continuity from 
small to large-angle and from LEP1 to LEP2 energies. 

The program contains all the necessary corrections to provide results, 
for typical experimental event selections, with the precision required 
to usefully compare with experimental measurements and other theoretical 
results. 
We estimate a precision of 0.1-0.2\% at small-angle, in the angular range 
of the new generation of LEP luminometers, mainly due to the inaccuracy in
in the two hard photon emission contribution. 
At large-angle for both LEP1 and LEP2 energies, we estimate a precision of 
0.5\% with the exception of the region of the beam energies a few GeV above 
the $Z$ boson resonance, where the relevance of the still missing calculation 
of the leading \Ord{\alpha^2 L^2} radiative corrections to the $Z-\gamma$ 
interference term spoils the accuracy up to 1\%, as in every other present 
calculation.

We have performed a detailed comparison of the photonic and weak corrections 
for both small-angle (angular range of the new generation of LEP luminometers) 
and large-angle Bhabha scattering. 

\begin{figure}[!ht]
\centering
\setlength{\unitlength}{0.1mm}
\begin{picture}(1700,1700)
\put( 400, 850){\makebox(0,0)[lb]{
\epsfig{file=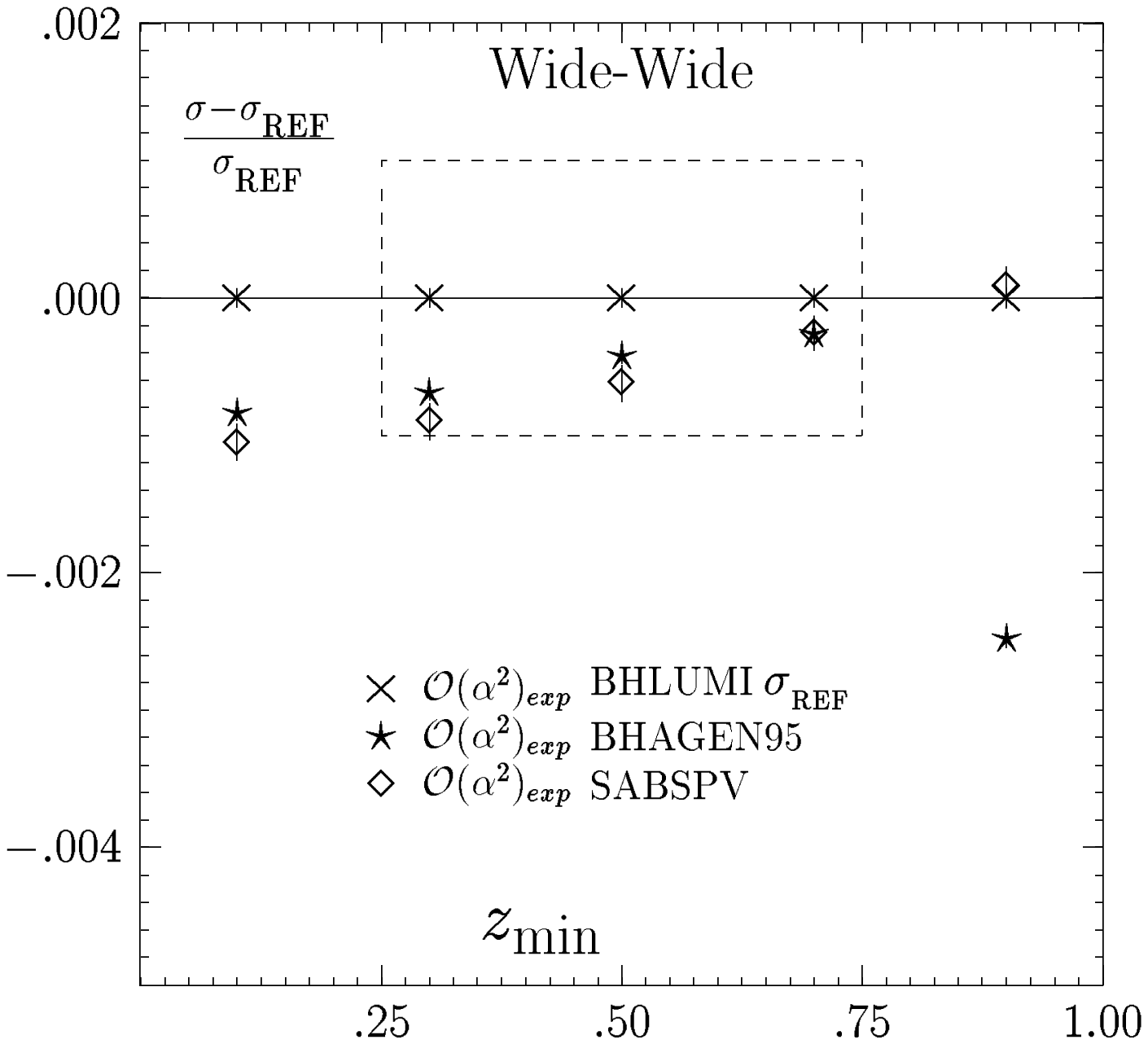,width=87mm,height=84mm}
}}
\put(-20, 00){\makebox(0,0)[lb]{
\epsfig{file=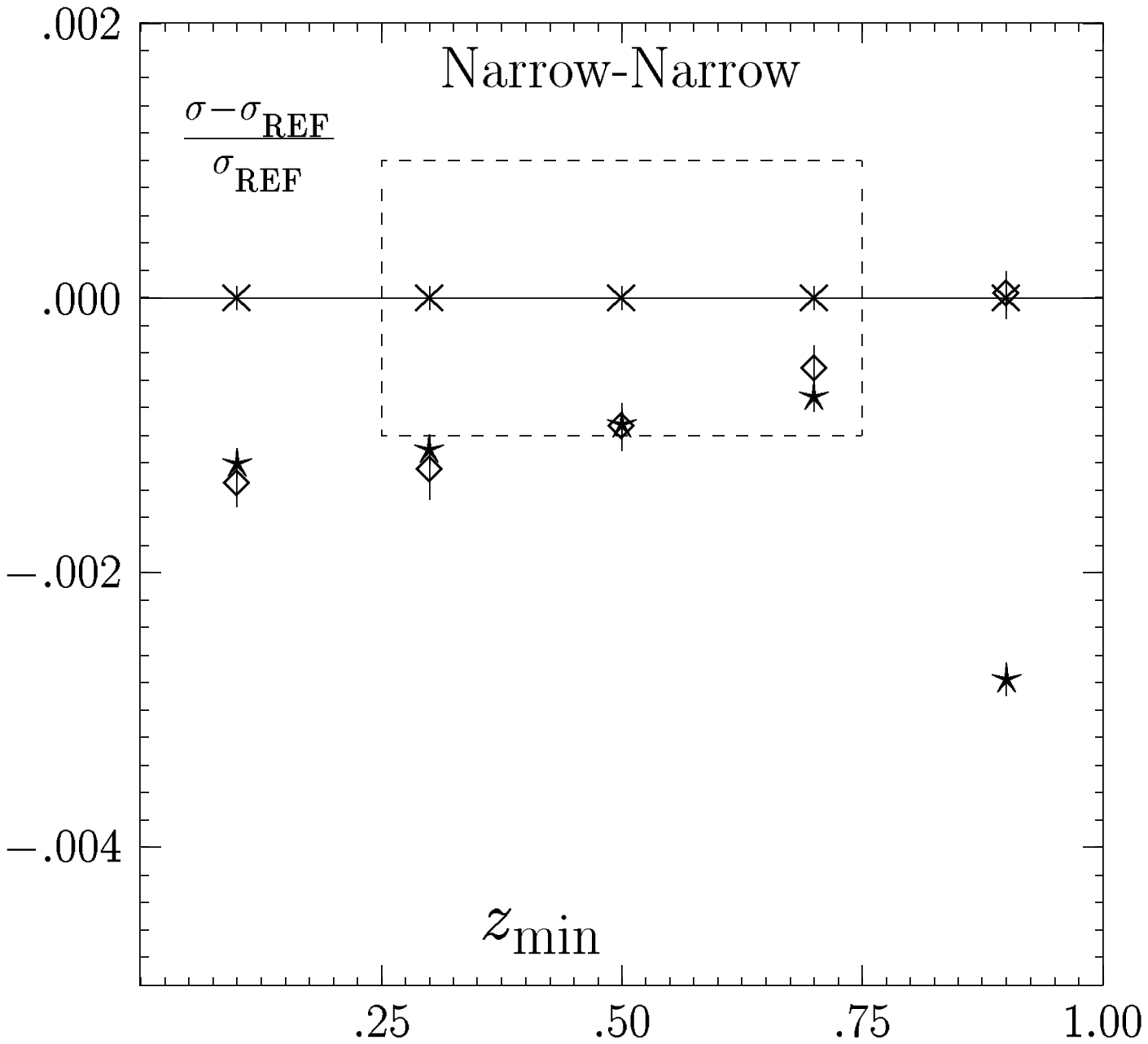,width=87mm,height=84mm}
}}
\put(820, 00){\makebox(0,0)[lb]{
\epsfig{file=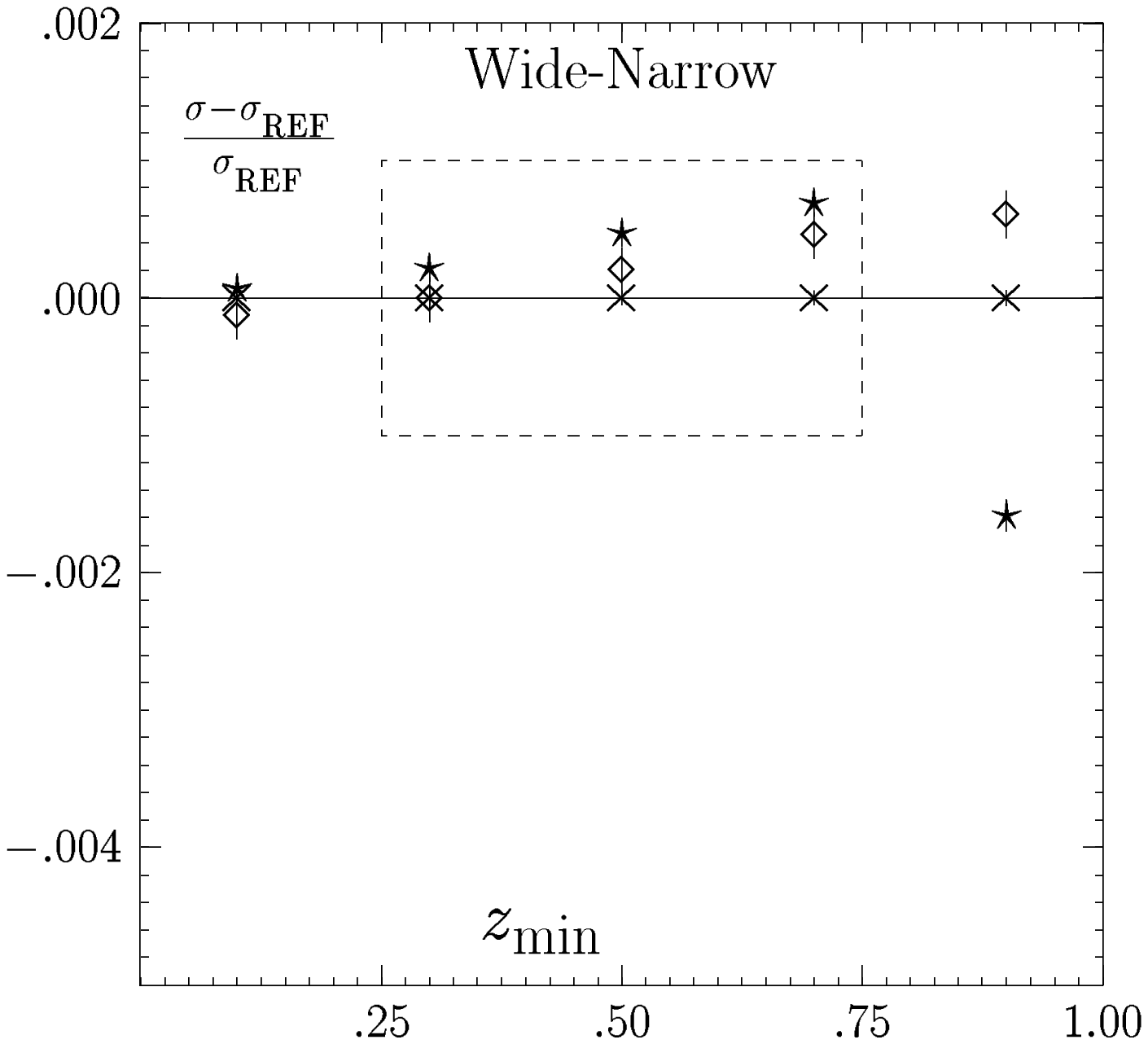,width=87mm,height=84mm}
}}
\put(0,0){\begin{picture}( 1600,1600)
  \put(  200,1200){\makebox(0,0)[b]{\huge All}}
  \put( 1500,1200){\makebox(0,0)[b]{\huge Included}}
\end{picture}} 
\end{picture}
\caption{\small\sf
 Monte Carlo results for various symmetric/asymmetric           
 versions of the CALO2 ES,
 for matrix elements beyond first order.                    
 Z exchange, up-down interference and vacuum polarization       
 are switched ON.                                               
 The center of mass energy is $\protect\sqrt{s}=92.3$~GeV.
 Not available x-sections are set to zero.                          
 In the plot, the ${\cal O}(\alpha^2)^{YFS}_{exp}$ cross section     
 $\sigma_{_{\rm{BHL}}}$ from BHLUMI 4.x                         
 is used as a reference cross section.                          
}
\label{fig:sical92asy-FG}
\end{figure}

At small-angle we obtain an agreement better than 0.02\% at \Ord{\alpha} 
with the program OLDBIS for every type of event selection. 
In the comparison beyond the \Ord{\alpha}, but limited to the $t$-channel only, 
with vacuum polarization and $Z-\gamma$ interference switched off, the results 
of BHAGEN95 agree with the ones of OLDBIS+LUMLOG better than 0.03\%, 
for typical experimental event selection, and have a difference less than 0.1\%
with BHLUMI and SABSPV in these conditions. 
For event selections less realistic for experiments the results of BHAGEN95
(and also of OLDBIS+LUMLOG) can differ from the ones of BHLUMI and SABSPV
of a few per mil. We believe to understand the difference (at least in the 
BHAGEN95 case) as due to the inaccuracy in the two hard photon emission 
contribution, which depends on the details of the event selection used and is 
smaller for looser cuts. 
When all possible contributions are switched on the agreement between 
BHLUMI, SABSPV and BHAGEN95 is inside 0.1\% for realistic experimental event
selection, as presented in Fig. \ref{fig:sical92asy-FG} (same as in Fig. 17 of 
~\cite{CERN96}). 
The better agreement in this case is due to a slightly larger (but less 
than 0.1\%) contribution of BHAGEN95, coming mainly from the implementation 
of the vacuum polarization and/or $Z-\gamma$ interference correction. 
We have investigated this small difference, but it is impossible to 
disentangle the contributions without ad hoc runs of all the compared programs.

\begin{figure}[!ht]
\centering
\setlength{\unitlength}{0.1mm}
\begin{picture}(1700,1800)
\put(-20, 00){\makebox(0,0)[lb]{
\epsfig{file=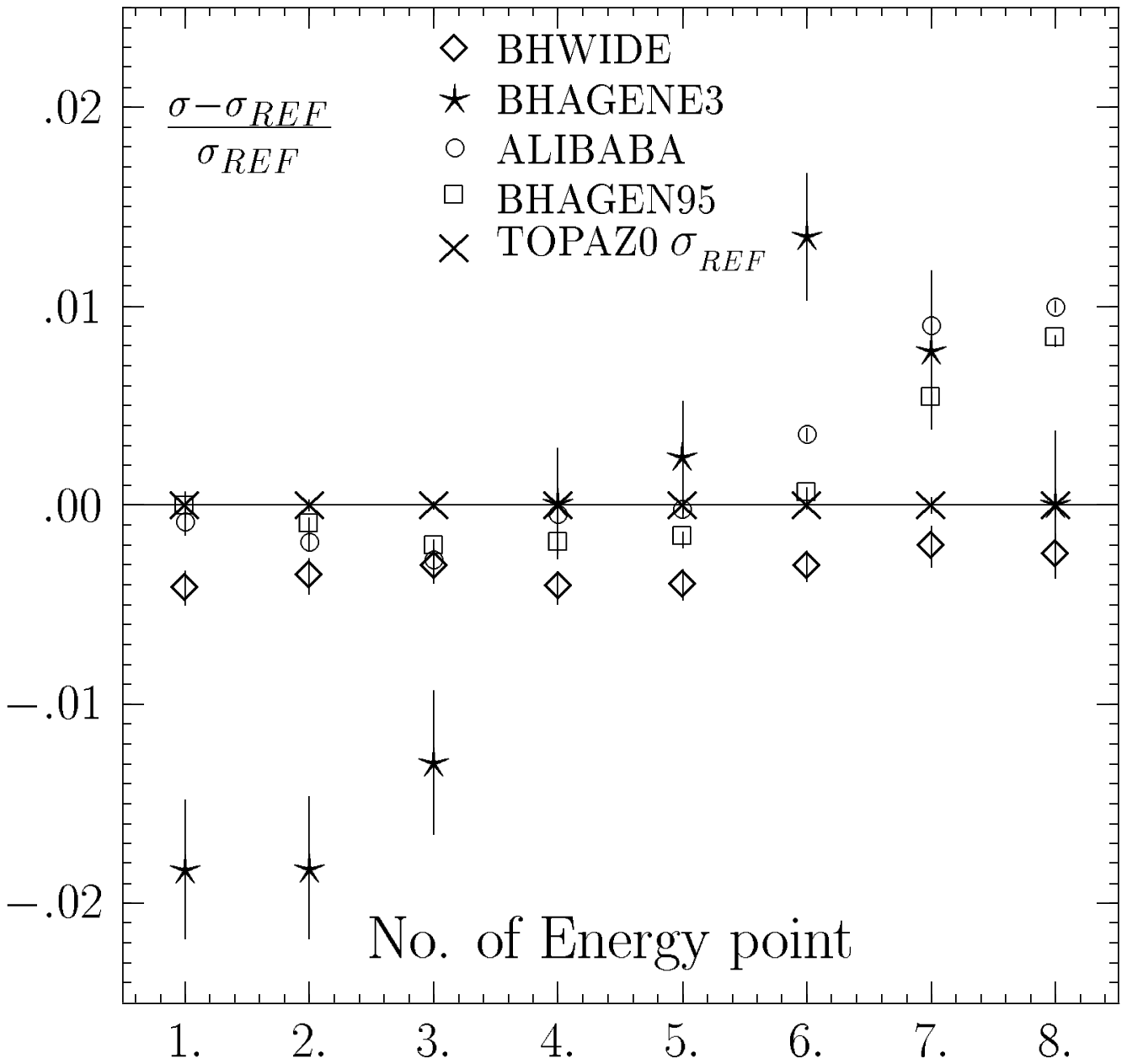,width=85mm,height=83mm}
}}
\put(830, 00){\makebox(0,0)[lb]{
\epsfig{file=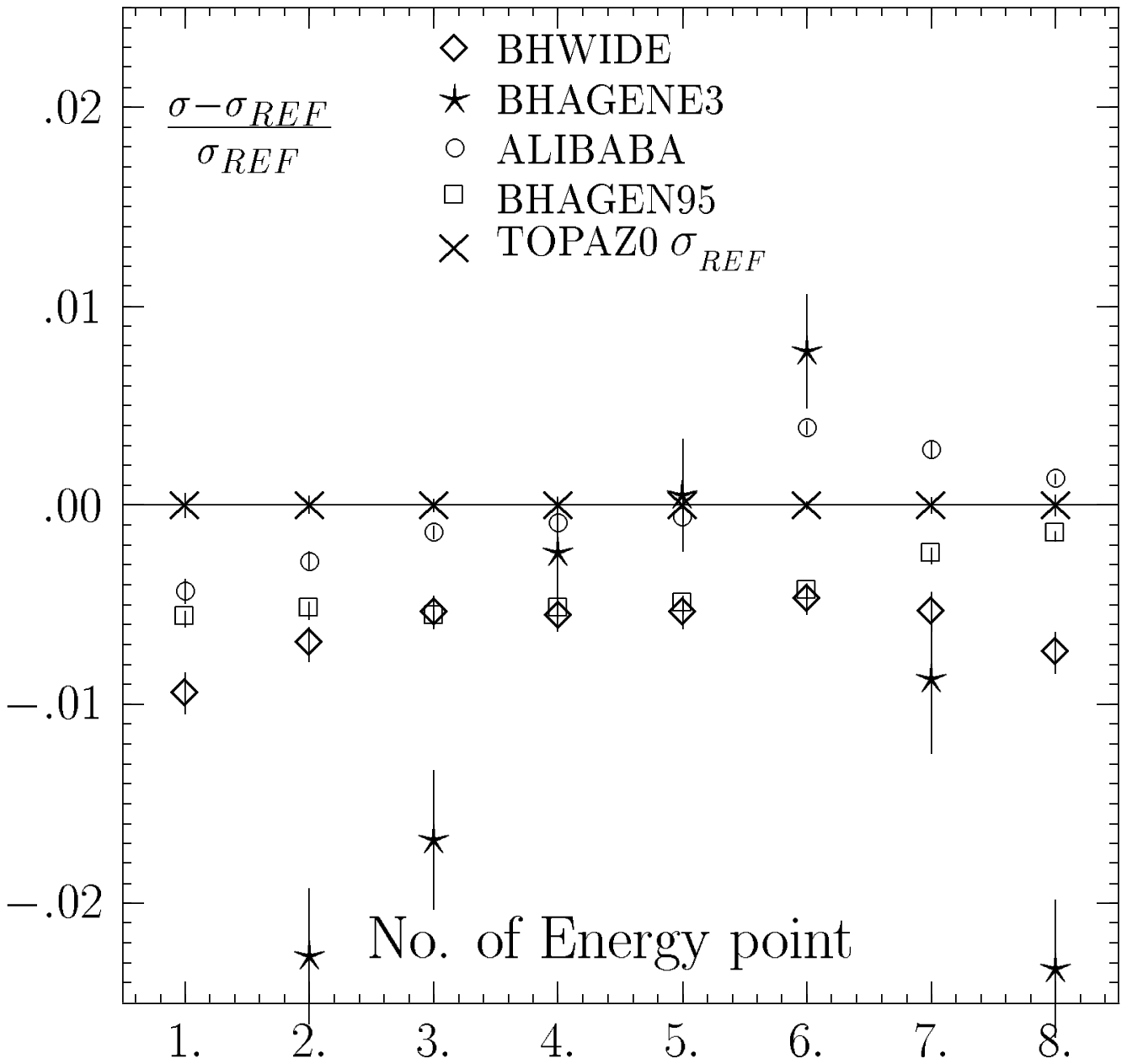,width=85mm,height=83mm}
}}
\put( 60,  900){\makebox(0,0)[lb]{
\epsfig{file=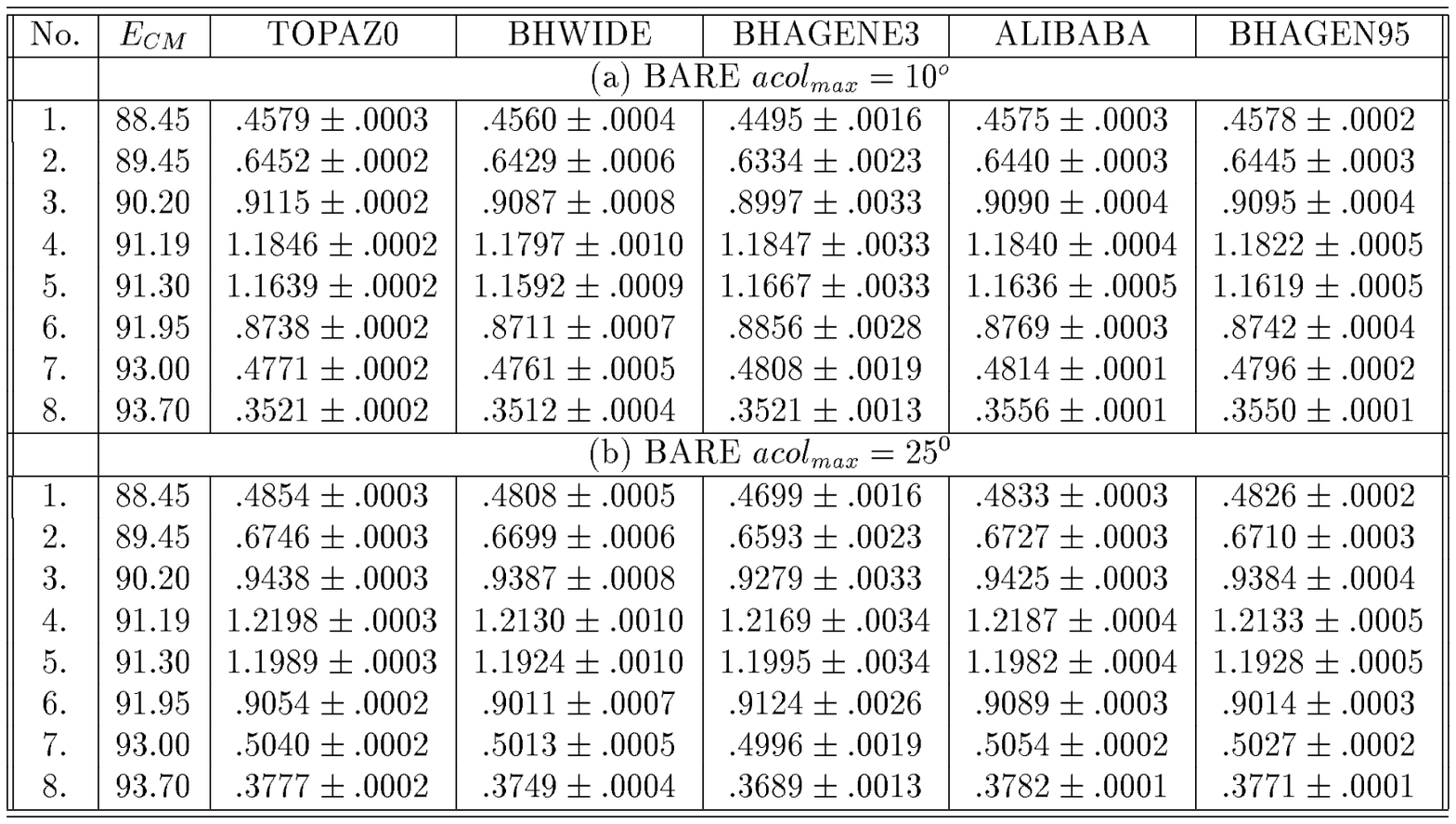,width=160mm,height=90mm}
}}
\put(0,0){\begin{picture}( 1600,1600)
  \put(  400, 850){\makebox(0,0)[b]{\small BARE $acol_{max}=10^o$ }}
  \put( 1200, 850){\makebox(0,0)[b]{\small BARE $acol_{max}=25^o$ }}
\end{picture}} 
\end{picture}
\caption{\small\sf
 Monte Carlo results for the BARE ES,
 for two values ($10^o$ and $25^o$) of acollinearity cut.
 Center of mass energies (in GeV) close to $Z$ peak.
 In the plots, the cross section
 $\sigma_{\rm{REF}}$ from TOPAZ0
 is used as a reference cross section. Cross sections in nb. 
}
\label{fig:lab-lep1-bare}
\end{figure}

\begin{figure}[!ht]
\centering
\setlength{\unitlength}{0.1mm}
\begin{picture}(1700,1800)
\put(-20, 00){\makebox(0,0)[lb]{
\epsfig{file=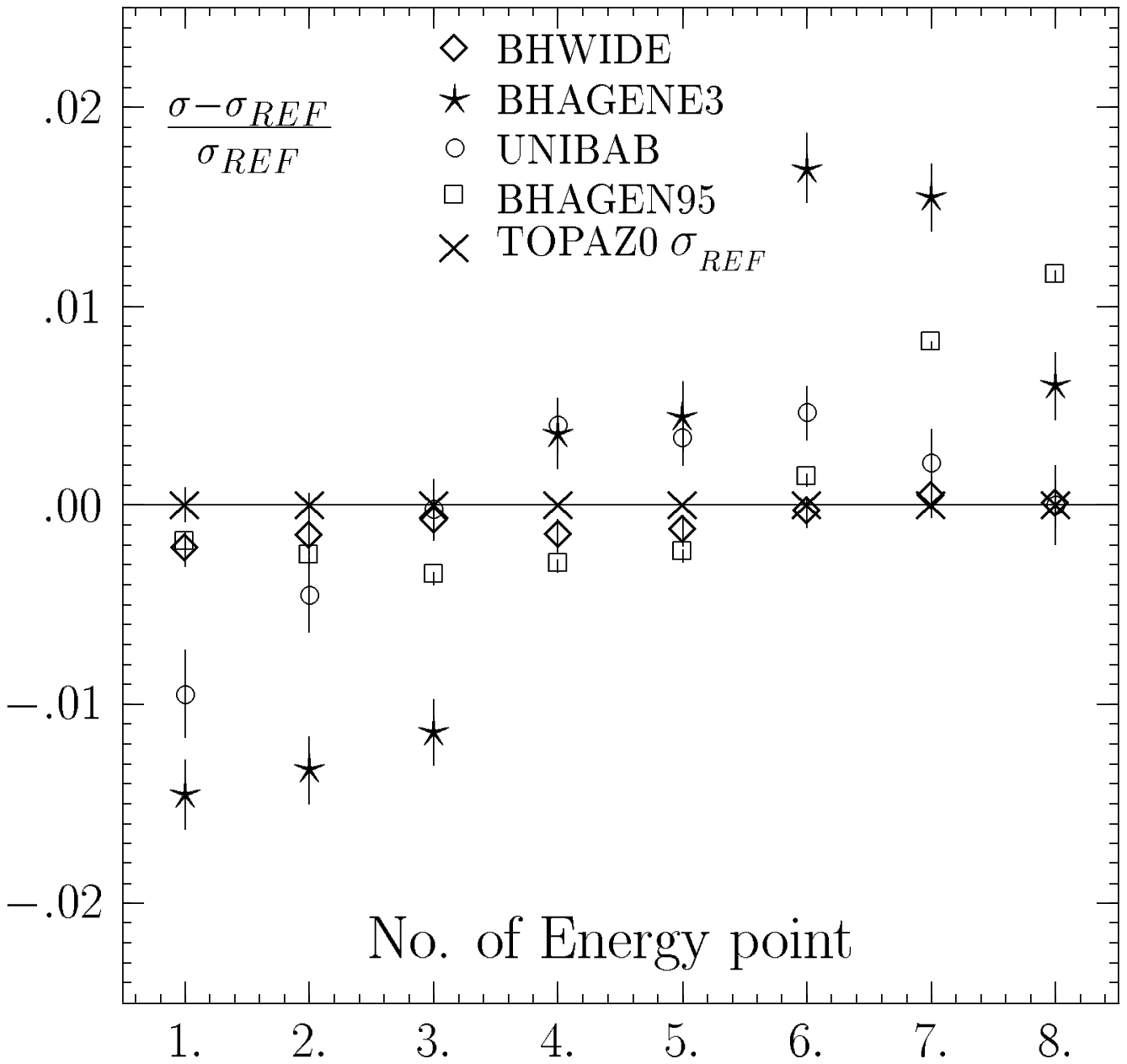,width=85mm,height=83mm}
}}
\put(830, 00){\makebox(0,0)[lb]{
\epsfig{file=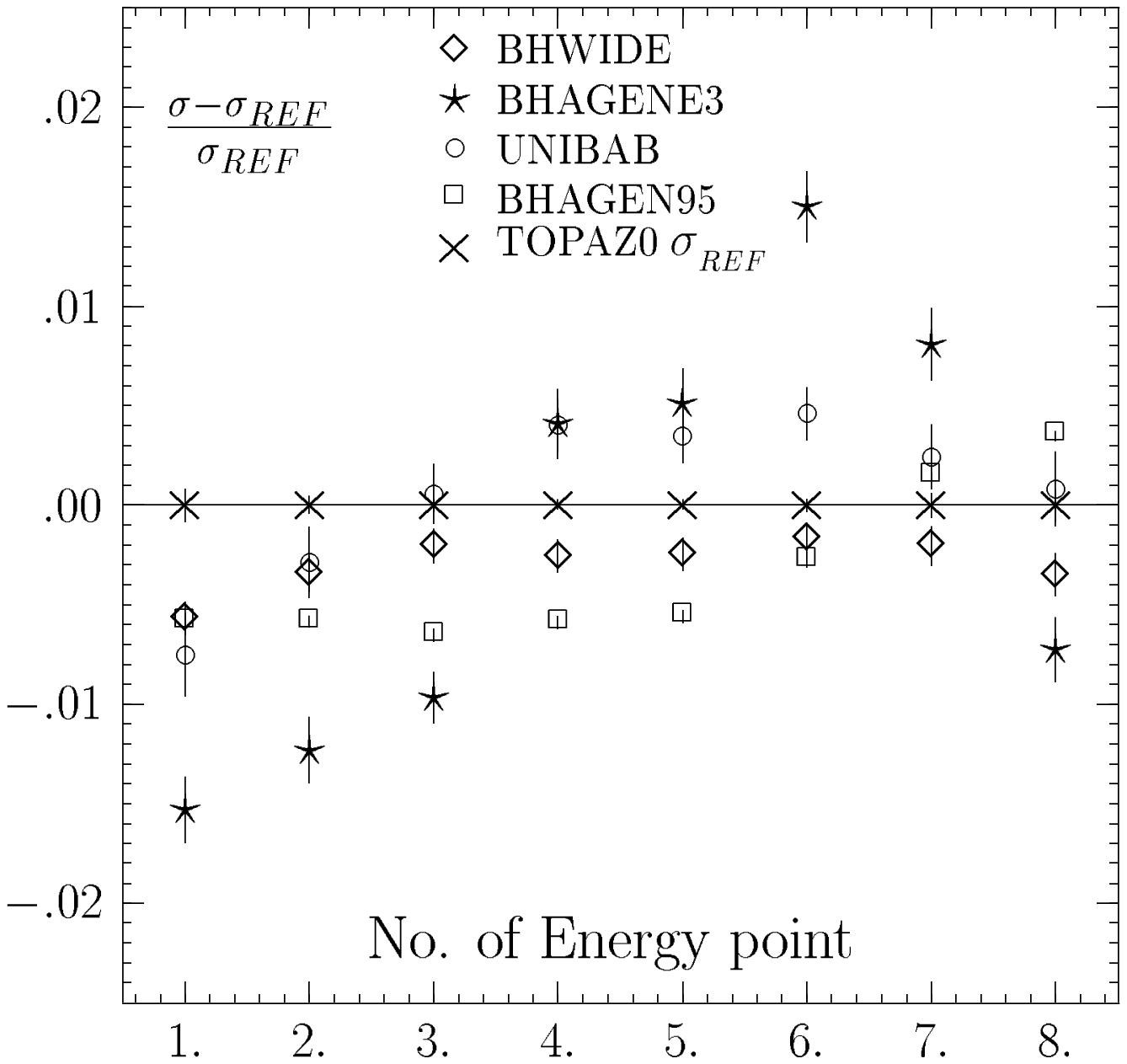,width=85mm,height=83mm}
}}
\put( 60,  900){\makebox(0,0)[lb]{
\epsfig{file=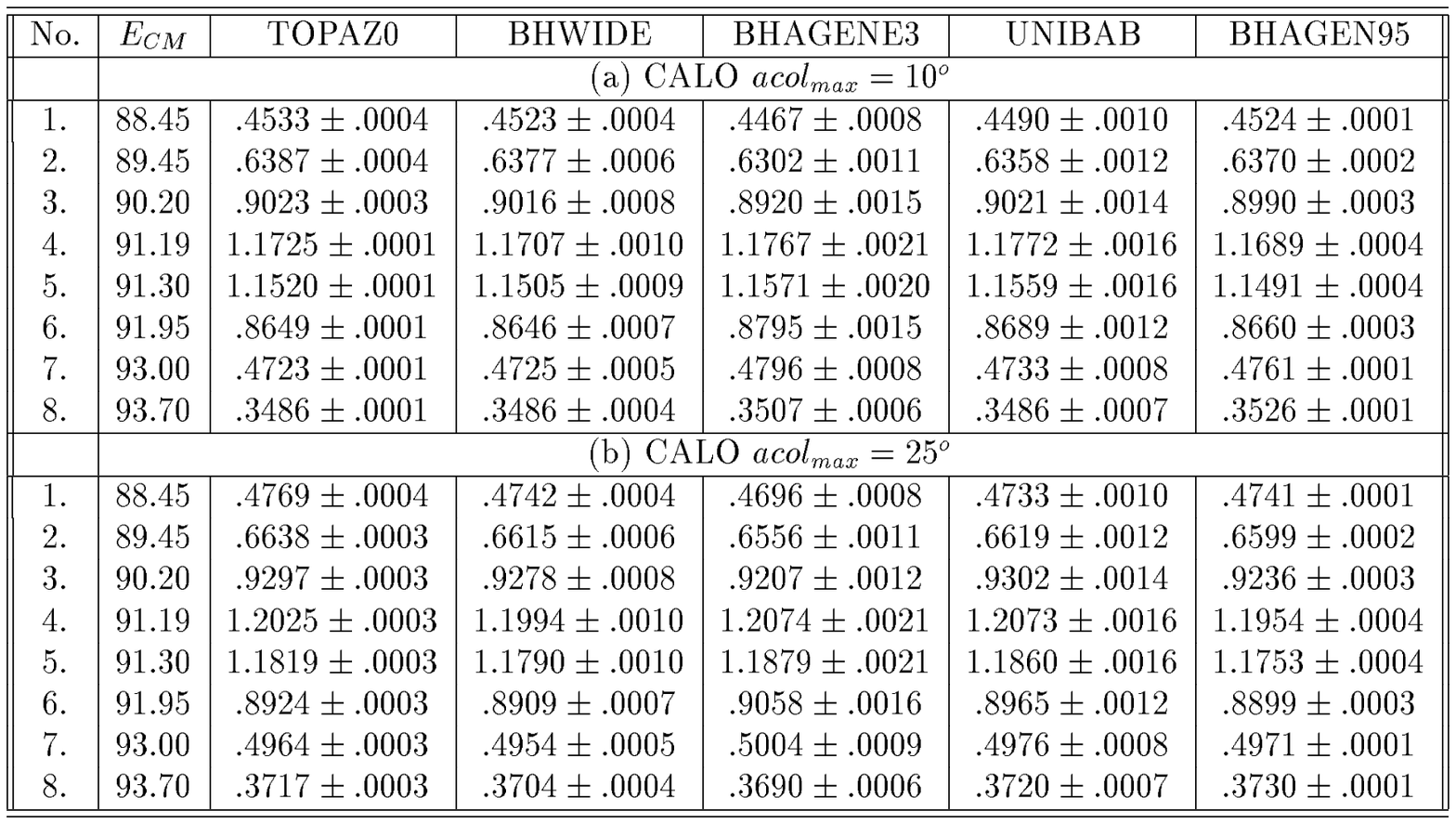,width=160mm,height=90mm}
}}
\put(0,0){\begin{picture}( 1600,1600)
  \put(  400, 850){\makebox(0,0)[b]{\small CALO $acol_{max}=10^o$ }}
  \put( 1200, 850){\makebox(0,0)[b]{\small CALO $acol_{max}=25^o$ }}
\end{picture}} 
\end{picture}
\caption{\small\sf
 Monte Carlo results for the CALO ES,
 for two values ($10^o$ and $25^o$) of acollinearity cut.
 Center of mass energies (in GeV) close to $Z$ peak.
 In the plots, the cross section
 $\sigma_{\rm{REF}}$ from TOPAZ0
 is used as a reference cross section. Cross sections in nb. 
}
\label{fig:lab-lep1-calo}
\end{figure}

At large-angle and in the LEP1 energy range the corrected results of BHAGEN95 
are shown in Fig. \ref{fig:lab-lep1-bare} and in Fig. \ref{fig:lab-lep1-calo}; 
the difference with the other results is 
within 0.5\% around the $Z$ boson resonance, being bigger 
for energies a few GeV above the peak, where BHAGEN95 is in agreement with
ALIBABA within 0.5\% and within 1\% with other codes.

\begin{figure}[!ht]
\centering
\setlength{\unitlength}{0.1mm}
\begin{picture}(1700,1400)
\put(-20, 00){\makebox(0,0)[lb]{
\epsfig{file=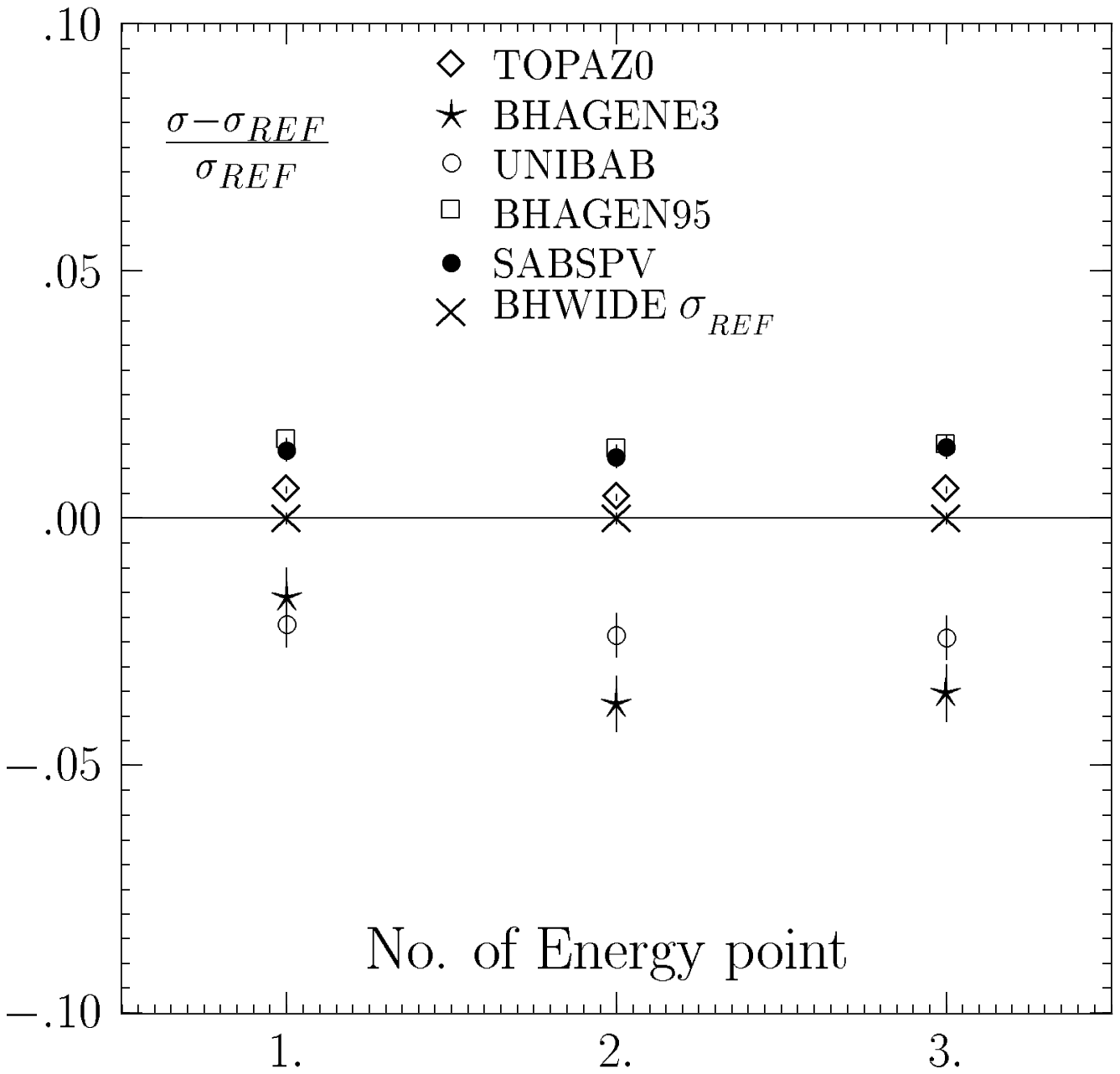,width=85mm,height=83mm}
}}
\put(830, 00){\makebox(0,0)[lb]{
\epsfig{file=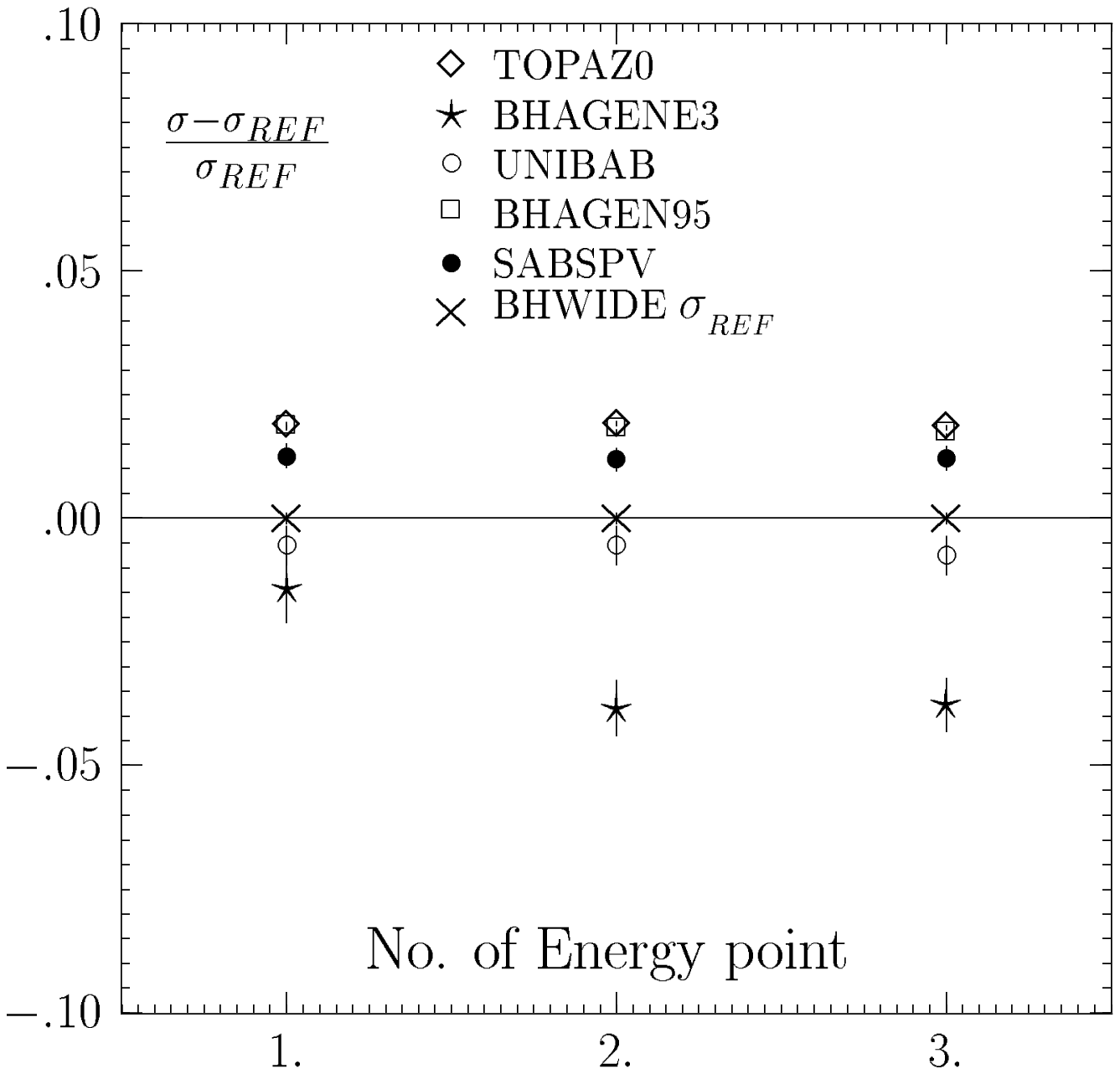,width=85mm,height=83mm}
}}
\put( 60,  900){\makebox(0,0)[lb]{
\epsfig{file=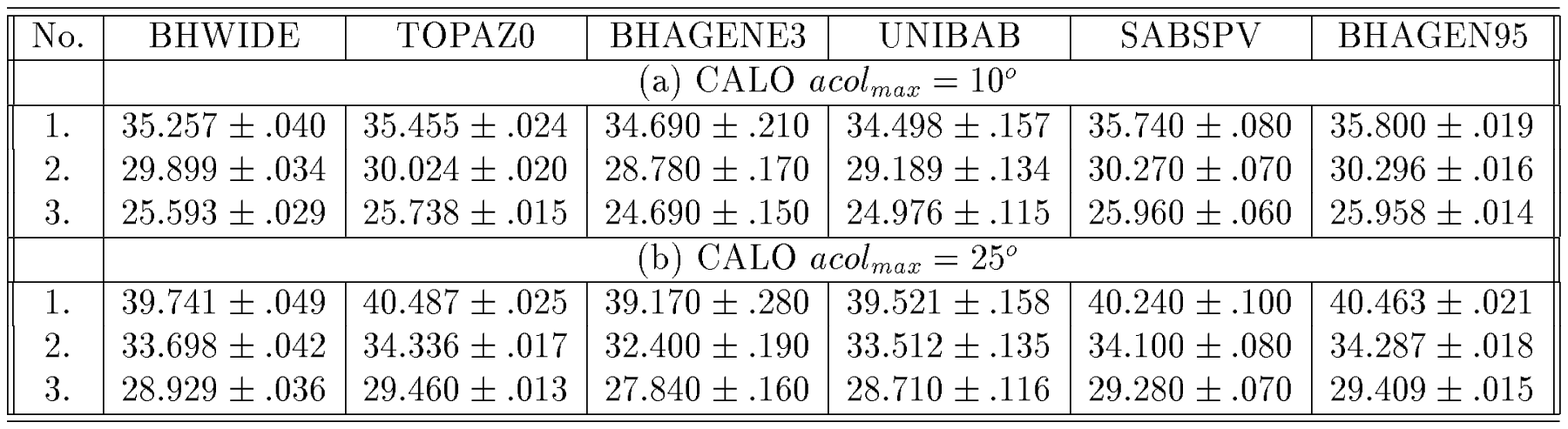,width=160mm,height=50mm}
}}
\put(0,0){\begin{picture}( 1600,1600)
  \put(  400, 850){\makebox(0,0)[b]{\small CALO $acol_{max}=10^o$ }}
  \put( 1200, 850){\makebox(0,0)[b]{\small CALO $acol_{max}=25^o$ }}
\end{picture}} 
\end{picture}
\caption{\small\sf
 Monte Carlo results for the CALO ES,
 for two values ($10^o$ and $25^o$) of acollinearity cut.
 Center of mass energies close to $W$-pair production threshold
 ($E_{CM}$: 1.~175~GeV, 2.~190~GeV, 3.~205~GeV).
 In the plots, the cross section
 $\sigma_{\rm{REF}}$ from BHWIDE
 is used as a reference cross section. Cross sections in pb. 
}
\label{fig:lab-lep2-calo}
\end{figure}

For LEP2 energies although the differences between the codes are larger 
(up to 2\% for the cluster of the results of TOPAZ0, BHWIDE, SABSPV and 
BHAGEN95), as shown in Fig. \ref{fig:lab-lep2-calo}, the precision is 
comparable with the foreseen experimental accuracy.
\bigskip

\noindent{\bf Acknowledgments}\par
Useful conversations with the conveners (S.~Jadach and O.~Nicrosini) and 
all the members of the CERN Working Group of the Event Generators for Bhabha 
Scattering are gratefully acknowledged. 
We thank W.~Hollik for useful informations about the BHM/WOH approach and 
W.~Beenakker about the ALIBABA parameters.
One of us (HC) is grateful to the Bologna Section of INFN and to the Department 
of Physics of Bologna University for support and kind hospitality.
\bigskip


\vfill \eject
\textwidth  = 17cm

%
\def\half{ {1\over 2} }
\def\alf1{ {\alpha\over\pi} }
%
%
%
%
\vfill \eject
%
%
%
%
\vfill \eject
%
%
%
%
\vfill \eject
%
%
%
%
\end{document}